\documentclass[11pt,british,english,11pt,a4paper]{article}
\usepackage[T1]{fontenc}
\usepackage[latin9]{inputenc}
\usepackage{verbatim}
\usepackage{textcomp}
\usepackage{amsmath}
\usepackage{amssymb}
\pdfoutput=1
\usepackage{graphicx}
\usepackage{esint}

\makeatletter

\newcommand{\lyxmathsym}[1]{\ifmmode\begingroup\def\b@ld{bold}
  \text{\ifx\math@version\b@ld\bfseries\fi#1}\endgroup\else#1\fi}

\usepackage{jcappub}

\usepackage{babel}

\usepackage{babel}
\usepackage{jcappub}

\makeatother

\usepackage{babel}
\begin{document}

\title{Interacting dark energy collapse with matter components separation}

\author[a,b]{M. Le Delliou}

\author[c,d]{and T. Barreiro}

\affiliation[a]{Departamento de Física Matemática, Instituto de Física, Universidade
de São Paulo,\\
CP 66.318 \textemdash{} 05314-970, São Paulo, SP, Brasil}

\affiliation[b]{Centro de Astronomia e Astrofísica da Universidade de Lisboa,\\
 Faculdade de Ciências, Ed. C8, Campo Grande, 1769-016 Lisboa, Portugal}

\affiliation[c]{Departamento de Matemática da FFMCC, Universidade Lusófona de Humanidades
e Tecnologias, \\
 Campo Grande, 376 \textemdash{} 1749-024 Lisboa, Portugal}

\affiliation[d]{Instituto de Plasmas e Fusão Nuclear, Instituto Superior Técnico,
\\
Avenida Rovisco Pais, 1 \textemdash{} 1049-001 Lisboa, Portugal}

\emailAdd{Morgan.LeDelliou@uam.es, delliou@cii.fc.ul.pt, tmbarreiro@ulusofona.pt}

\abstract{We use the spherical collapse model of structure formation to investigate
the separation in the collapse of uncoupled matter (essentially baryons)
and coupled dark matter in an interacting dark energy scenario. Following
the usual assumption of a single radius of collapse for all species,
we show that we only need to evolve the uncoupled matter sector to
obtain the evolution for all matter components. This gives us more
information on the collapse with a simplified set of evolution equations
compared with the usual approaches. We then apply these results to
four quintessence potentials and show how we can discriminate between
different quintessence models.}


\arxivnumber{0812.6373 }

\date{submitted: 29/08/2012 -- accepted 31/01/2013}

\maketitle

\section{Introduction}

The field of cosmology, although receiving an accumulation of precision
observations in the recent years and despite their consistence with
the {}``concordance model'', $\Lambda$CDM, is yet to decide on
the content of the universe. The wealth of data available ranges from
large scale geometric assessments like the precise measurements of
the Cosmic Microwave Background (CMB) \cite{Komatsu:2010fb}, intermediate
scale evaluations like Baryon Acoustic Oscillations (BAO) \cite{Lampeitl:2009jq}
and more local scales measures of the Hubble constant ($H_{0}$) via
observations of standard candles like Cepheid variable stars \cite{Riess:2009pu}
or Type Ia Supernovae (SNIa) \cite{Kowalski:2008ez}. Although the
simplicity of the concordance model would place it in a prominent
position, given its compatibility with observations, it suffers from
theoretical shortcomings that question our understanding of gravity
and our knowledge of physics: the fine tuning problem and the coincidence
problem (for a review, see \cite{Carroll:2000fy}). Introduction of
dynamical forms of dark energy (hereafter DE) was then considered
to try and overcome those problems \cite{Carroll:2000fy,Sahni:1999gb,Padmanabhan:2004av},
and quintessence scalar fields \cite{Wetterich:1987fm,Ratra:1987rm,Wetterich:1994bg}
are the most obvious candidates (for review see \cite{Copeland:2006wr}).
However if they can overcome the first problem, the second remains
in various forms. In this framework, DE and dark matter (DM) are the
fundamental building blocks of the cosmological standard model based
on general relativity, which leaves unknown and dark nearly 96\% of
the Universe's energy content. Although coupling to baryonic matter
is tightly constrained by local gravity tests \cite{Carroll:1998zi},
this undetermined aspect opens the door to all kinds of couplings
within the dark sector, including non-linear DE self-gravitating clustering
\cite{Mota:2004pa,Nunes:2004wn} and non-minimal coupling between
the two dark species, introduced either by ad hoc arguments \cite{Amendola:1999er}
or because dark energy and dark matter are unified in the context
of some framework \cite{Kamenshchik:2001cp,Bento:2002ps,Bertolami:2007wb}:
the implications of interacting DE-DM on cosmological observations
were examined in works such as \cite[and references therein]{Amendola:1999er,Amendola:2003wa,Pettorino:2008ez,Brookfield:2007au,Rosenfeld:2007ri},
while analysis of CMB and BAO such as \cite{Amendola:2006dg,Bean:2008ac},
or of gamma-ray bursts as in \cite{Barreiro:2010nb}, have constrained
some coupled quintessence models. Studies have demonstrated that signatures
of such coupled models can be found in the background expansion history
of the universe and on cosmic structure formation, especially at larger,
galaxy clusters, scales \cite{Baldi:2008ay,Baldi:2010vv,Mainini:2006zj,Sutter:2008fs,Maccio:2003yk}.
However, despite the naturalness of allowing dark sector interaction,
a degeneracy exists between the various possible amounts of interacting
DM at the homogeneous dynamical level \cite{Kunz:2007rk}, that only
non-linear clustering properties might disentangle. Another consequence
of interacting dark sector could be the violation of the Equivalence
Principle: originating in the DE-DM interactions \cite{Kesden:2006vz,Kesden:2006zb,Farrar:2006tb,Bertolami:2007zm,Bertolami:2007tq,Bertolami:2012yp,Abdalla:2007rd,Warnick:2008tk,Abdalla:2009mt,Peebles:2009th,Sanderson:2010ct},
this could be seen in differential DM-baryons tidal arms from disrupted
satellite galaxies \cite{Kesden:2006vz,Kesden:2006zb,Warnick:2008tk,Peebles:2009th},
totally disrupted satellites like the bullet cluster \cite{Farrar:2006tb}
or DM caustics \cite{Sanderson:2010ct}.

Dynamic DE impact on structure formation, in the form of a quintessence
scalar field, was first restricted to linear or perturbation theory
of structure evolution \cite{Benabed:2001dm}. In the mildly non-linear
regime, \cite{Ma:1999dwa,Lokas:2000cn} found the smallest scale of
quintessence fluctuations are larger than the clusters scale, so this
clustering should be negligible. Nevertheless, studies of the fully
non-linear regime showed marked differences in DM structures between
DE models that cluster and those that do not \cite{Mota:2004pa}.
The clustering properties of DE models remain however an open question.
Non-linear structure formation is traditionally studied with N-body
simulations, as e.g. in \cite{BoylanKolchin:2009nc,Marulli:2011jk}.
Insofar as investigating structure formation with DE, since no dynamical
model seem preferred, the flexibility of semi-analytical methods have
lead to numerous studies such as \cite{Lokas:2000cn,Mainini:2002ry,Mainini:2003uf,Mota:2004pa,Lokas:2003cj,Solevi:2004tk,Solevi:2005ta,Nunes:2004wn,Manera:2005ct,LeDelliou:2005ig,Tarrant:2011qe}.
The methods used, prevalent at the core of entirely semi-analytic
galaxy formation models, i.e in \cite{Kauffmann:1994gv,Baugh:1997nc,Somerville:1998bb,Cole:2000ex,Hatton:2003du,LeDelliou:2004wh,LeDelliou:2005hj},
revolve around similar mass function-determining schemes as the methodology
developed by Press \& Schechter \cite[hereafter PS]{Press:1973iz}.
The latter method uses a spherically symmetric dynamical model to
relate the collapse of massive structures to a density threshold in
the linearly extrapolated density field. In this way, it is possible
to apply Gaussian statistics to the initial density field in order
to count the numbers of collapsed structures above a given mass threshold
at a particular epoch. The spherical collapse model \cite{Larson:1969mx,Penston:1969yy,Gunn:1972sv,Fillmore:1984wk,Bertschinger:1900nj,Peebles80,LeDPhD,Henriksen:2002nt,LeDelliou:2003mk,Delliou:2007hu,Henriksen:2009be,Henriksen:2009bh,Henriksen:2009bd}
or its variants are therefore at the root of such approach.

Non-gravitational interactions is expected to impact significantly
on the free fall of DM, whether it links baryons with DM \cite{Stubbs:1993xk}
or originates in interacting DE \cite{Kesden:2006vz,Kesden:2006zb,Farrar:2006tb,Warnick:2008tk,Peebles:2009th,Sanderson:2010ct}.
This investigation aims at revealing that impact in the spherical
collapse model, as suggested in \cite{Bertolami:2007zm,Bertolami:2007tq},
in its dependence on the DE model involved and on the difference between
the collapses of coupled and uncoupled matter species, as mark of
a baryon/DM segregation. For that purpose, we singled out a range
of quintessence potentials and a range of interaction parameters to
gauge their influence on the violation of the Equivalence Principle
between coupled and uncoupled matter manifested by this segregation:
the double exponential potential \cite{Barreiro:1999zs}, the infinite
sum of inverse power potential \cite{Steinhardt:1999nw}, the supergravity
motivated potential (SUGRA) \cite{Brax:1999yv}, and the superstring
theory motivated potential of \cite{Albrecht:1999rm}. For these models,
we used the spherical collapse model to obtain separately the uncoupled
matter  and DM overdensity parameters evolution and their critical
values, arguably showing how the baryon/DM drift can manifest itself
depending on the DE model. We introduced a novel approach, noticing
that a single collapse radius kept for all species lead to algebraic
relations between their overdensities. Thus we simplified the treatment
of dynamics which entails following dust-like uncoupled matter in
a DE environment. This simpler evolution manages to recover previous
results such as \cite{Nunes:2004wn,Tarrant:2011qe} and in the process
allows to differentiate between coupled and uncoupled matter evolution.

The paper is organised as follows: in section   \ref{sec:The-spherical-collapse},
the evolution equations for both the baryons or uncoupled matter and
for the coupled DM are detailed, followed by the forms of the potentials;
section   \ref{sec:Numerical-implementation-and} presents our analysis
of the segregated spherical collapse semi-analytical model. Finally,
in section   \ref{sec:Discussion-and-conclusion} we discuss the results
and present our conclusions.

\section{The spherical collapse in interacting DM/DE quintessence models\label{sec:The-spherical-collapse}}

The spherical collapse model (pioneered in \cite{Larson:1969mx,Penston:1969yy,Gunn:1972sv,Fillmore:1984wk,Bertschinger:1900nj}
and summarised in \cite{Peebles80}) is a powerful tool of semianalytical
methods to study gravitational clustering, e.g. \cite{LeDPhD,Henriksen:2002nt,LeDelliou:2003mk,Delliou:2007hu,Henriksen:2009be,Henriksen:2009bh,Henriksen:2009bd},
in particular with many different models of interacting DE as in \cite{Lokas:2000cn,Mainini:2002ry,Mainini:2003uf,Mota:2004pa,Lokas:2003cj,Solevi:2004tk,Solevi:2005ta,Nunes:2004wn,Manera:2005ct,LeDelliou:2005ig,Tarrant:2011qe}.

The DE models considered here are interacting quintessence dynamical
scalar fields. The governing equations of motion come from Einstein's
field equations (Friedmann and Raychaudhuri equations) and Bianchi
identities (fluids energy density conservations), for a multicomponent
fluid with a top hat density model. The top hat model assumes a Friedmann-Lemaître-Robertson-Walker
(FLRW) background and a higher curved FLRW spherical collapsing patch.
In this paper we will always assume the background to be flat (no
curvature). The multicomponent fluid comprises baryons, uncoupled
DM, coupled DM and DE. We treat the baryons and uncoupled DM in a
single uncoupled matter fluid. The Bianchi identities for each species
reflect the coupling of DE as well as its clustering properties in
the collapsing patch with heat fluxes. The scalar field obeys a corresponding
Klein-Gordon equation. The type of quintessence is entirely determined
by its potential, however an important part of the model, as far as
structure formation is concerned, is given in the interaction parameter
as well as the clustering parameter.

\subsection{Background evolution\label{sub:Background-evolution}}

we will use the following notations 
\begin{itemize}
\item energy densities:

$\rho_{\phi}$ for DE energy density (quintessence from a scalar field
$\phi$), $\rho_{m}$ for total matter, $\rho_{u}$ for uncoupled
matter (including uncoupled DM and baryons) and $\rho_{c}$ for coupled
DM. These energy densities are related by

\begin{align}
\rho_{m}= & \rho_{u}+\rho_{c},
\end{align}
where the uncoupled matter component includes the baryon and the uncoupled
dark matter energy components, 
\begin{align}
\rho_{u}= & \rho_{b}+\rho_{udm}.
\end{align}
For the total energy density, we use 
\begin{align}
\rho_{tot} & =\rho_{m}+\rho_{\phi}.
\end{align}

\item density parameters

we define the density parameter for each species as
\begin{align}
\Omega_{i}= & \frac{\rho_{i}}{\rho_{tot}}.
\end{align}

\item FLRW Hubble parameter defined in terms of scale factor $a$:

\begin{align}
H= & \frac{1}{a}\frac{da}{dt}=\frac{\dot{a}}{a},
\end{align}

\end{itemize}

\subsubsection{Background Friedmann evolution}

The Einstein field equations for the flat FLRW background are reduced
to the Friedmann and Raychaudhuri equations for the background scale
factor $a$ 
\begin{align}
H^{2}= & \frac{\kappa^{2}}{3}\left(\rho_{m}+\rho_{\phi}\right) & \dot{H}+H^{2}= & -\frac{\kappa^{2}}{6}\left(\rho_{m}+\rho_{\phi}+3P_{\phi}\right),\label{eq:FriedmannRaychadhuri}
\end{align}
with $P_{\phi}$ being the pressure of the dark energy.

\subsubsection{Energy density conservation equations}

Aside from the Einstein equations, the conservation of each cosmic
fluid component is governed by the corresponding Bianchi identity
for the homogeneous FLRW model, so we have for the uncoupled, coupled
matter and DE 
\begin{align}
\dot{\rho}_{u}+3H\rho_{u}= & 0, & \dot{\rho}_{c}+3H\rho_{c}= & \rho_{c}B^{\prime}\dot{\phi}, & \dot{\rho}_{\phi}+3H\left(\rho_{\phi}+P_{\phi}\right)= & -\rho_{c}B^{\prime}\dot{\phi},\label{eq:bkgU+C+DE}
\end{align}
where the DE-DM interaction manifests in the heat fluxes $\Gamma_{c}=-\Gamma_{\phi}=\rho_{c}B^{\prime}\dot{\phi}$,
with $B^{\prime}=\frac{dB(\phi)}{d\phi}$ the interaction rate. The
first proposals for such a flux appeared in \cite{Amendola:1999er}
with the choice of a constant $B^{\prime}$. Other forms can be chosen,
as, e.g., in \cite{Mainini:2004he}, however we will concentrate here
on the form $B^{\prime}=-C\kappa$ with the free parameter $C$ chosen
appropriately ($C=0$ representing of course absence of DE-DM coupling).
Both matter conservation equations (\ref{eq:bkgU+C+DE}-a,b) can be
integrated, yielding 
\begin{align}
\rho_{u}= & \rho_{0}\Omega_{u0}\left(\frac{a_{0}}{a}\right)^{3}, & \rho_{c}= & \rho_{0}\Omega_{c0}\left(\frac{a_{0}}{a}\right)^{3}e^{B-B_{0}},\label{eq:rhoUbCb}
\end{align}
where $\rho_{0}$ is the critical density at present, the subscript
0 denoting quantities at present, so $B_{0}=B(\phi_{0})$. Note that
$e^{B}$ can be interpreted as the mass of the coupled DM.

\subsection{Choice of potential\label{sub:Choice-of-potential}}

We chose a range of potentials $V(\phi)$ likely to reflect a variety
of DE behaviours. The double exponential potential is essentially
a scaling model, like a single exponential potential , that is complemented
by the second exponential to allow the scalar field to start dominating
the energy density at present times. The Albrecht-Skordis potential
is the only potential we use that has a local minimum, so that the
present energy density for the scalar field is obtained through oscillations
around this minimum. The Steinhardt and the SUGRA potentials are both
typical tracking potentials. In appendix~\ref{sec:observations}
we use the present SNeIa and CMB observations to constraint the coupling
$C$ we use in these models, where we can verify that the Steinhardt
and the SUGRA potentials favor a non-zero value for the coupling parameter
$C$, positive for the Steinhardt potential and negative for the SUGRA
potential.

In this framework, the energy density and pressure of the scalar field
are given by 
\begin{align}
\rho_{\phi} & =\frac{\dot{\phi}^{2}}{2}+V(\phi), & P_{\phi} & =\frac{\dot{\phi}^{2}}{2}-V(\phi).\label{eq:rhoQ,P_Q}
\end{align}
 They come into play in eq. (\ref{eq:bkgU+C+DE}-c) which is, for
a scalar field, the Klein-Gordon equation 
\begin{align}
\ddot{\phi}+3H\dot{\phi}+V^{\prime} & =-\rho_{c}B^{\prime}.\label{eq:KleinGordon}
\end{align}

\subsubsection{Double exponential}

The double exponential potential was introduced by \cite{Barreiro:1999zs}
to build a more realistic DE candidate out of the two regimes of the
single exponential potential \cite{Wetterich:1987fm,Ferreira:1997hj}:
one which tend to dominate the energy density, the other which mimics
the time evolution of the background. Originally, the simple exponential
potential gets its motivation from its display of a generic form for
moduli fields from extradimensional theories flat directions. Potentials
of this type could arise in string theory from Kaluza\textendash{}Klein
type compactification \cite{Barreiro:1999zs}. The result is called
double exponential potential (2EXP), which follows its name 
\begin{align}
V(\phi)= & \Lambda_{\phi}^{4}\left(e^{-\lambda_{1}\phi}+e^{-\lambda_{2}\phi}\right).\label{eq:2EXP}
\end{align}

Throughout the rest of the paper all numerical evolutions for this
potential are done with the parameter values $\lambda_{1}=20.5$ and
$\lambda_{2}=0.5$. The value of $\Lambda_{\phi}$ is fine-tuned to
achieve a present value of $\Omega_{\phi}=0.7$.

\subsubsection{Albrecht-Skordis}

The combination of a single exponential with a phenomenologically
inspired polynomial, motivated by superstring theory, introduced by
\cite{Albrecht:1999rm} creates a local minimum in which the field
can oscillate and get trapped. The parameters orders of magnitude
are then more natural ($\sim O(M_{Pl})$) for a range where the model
is able to satisfy a number of cosmological constraints and provide
late time acceleration, however they still require fine tuning caused
by the coincidence problem. It can take the form (AS) 
\begin{align}
V(\phi) & =\Lambda_{\phi}e^{-\kappa\lambda\phi}\left[A+\left(\kappa\phi\lyxmathsym{\textminus}B\right)^{2}\right].\label{eq:AS}
\end{align}

The values of the parameters we used for our numerical evolutions
were $A=0.01$, $B=33.96$ and $\lambda=8$. Again, $\Lambda_{\phi}$
is fine-tuned to obtain $\Omega_{\phi}=0.7$ at present.

\subsubsection{Steinhardt}

In \cite{Steinhardt:1999nw}, Steinhardt propose this new form of
potential by taking the infinite sum of inverse power potentials that
would stem from global SUSY considerations \cite{Binetruy:1998rz}.
It exemplifies well the tracking properties, desirable for quitessence
potentials to correlate the fine tuning problem to the coincidence
problem and thus reduce the theoretical cosmological constant problems
to one. It can be written as (SP) 
\begin{align}
V(\phi) & =\Lambda_{\phi}^{4}\left(e^{\frac{1}{\kappa\phi}}-1\right).\label{eq:SP}
\end{align}

The Steinhardt potential has no extra free parameters and $\Lambda_{\phi}$
is fine-tuned to achieve $\Omega_{\phi}=0.7$ at present.

\subsubsection{SUGRA}

\cite{Brax:1999yv} motivated this potential from supergravity models
and null vacuum superpotential expectation value in the low energy
approximation. It takes the form (SG) 
\begin{align}
V(\phi) & =\frac{\Lambda_{\phi}^{4+\alpha_{\phi}}}{\phi^{\alpha_{\phi}}}e^{\kappa^{2}\phi^{2}/2}.\label{eq:SUGRA}
\end{align}

For our numerical results we used $\alpha_{\phi}=11$, and fine-tuned
$\Lambda_{\phi}$ to achieve a value of $\Omega_{\phi}=0.7$ at present.

\subsection{Collapse evolution equations}

The aim of this study is to extend the usual spherical collapse to
follow separately the evolution of the coupled and uncoupled (including
baryons) components of matter. The homogeneous degeneracy between
those two components \cite{Kunz:2007rk} is expected to be broken
once we take into account their clustering properties. In the usual
scenario, a FLRW flat background contains a central spherical patch,
modeled as a positively curved FLRW with scale factor $r$. That scale
factor is assimilated to the radius of the spherical patch \cite{Lokas:2000cn,Mainini:2002ry,Mainini:2003uf,Mota:2004pa,Lokas:2003cj,Solevi:2004tk,Solevi:2005ta,Nunes:2004wn,Manera:2005ct,LeDelliou:2005ig,Tarrant:2011qe}.

\subsubsection{Uncoupled collapse}

The collapsing uncoupled matter in this framework follows the conservation
equation using the Hubble parameter $\mathcal{H}=\frac{\dot{r}}{r}$
in the collapsed patch, 
\begin{align}
\dot{\rho}_{u\star}+3\mathcal{H}\rho_{u\star}= & 0,\label{eq:CollU}
\end{align}
 where henceforth we will denote with a $\star$ subscript all collapsing
quantities. This integrates to 
\begin{align}
\rho_{u\star}= & \rho_{u\star i}\left(\frac{r_{i}}{r}\right)^{3},
\end{align}
 where the $i$ subscript denotes values at some arbitrary initial
time of collapse. Therefore the ratio of the uncoupled collapsed matter
to its background (\ref{eq:rhoUbCb}-a) is then related to the radius
through 
\begin{align}
\frac{\rho_{u\star}}{\rho_{u}}\propto & \left(\frac{a}{r}\right)^{3}.\label{eq:Uratio}
\end{align}

\subsubsection{Coupled collapse}

On the same token, the collapsing coupled DM follows the conservation
equation 
\begin{align}
\dot{\rho}_{c\star}+3\mathcal{H}\rho_{c\star}= & \rho_{c\star}B_{\star}^{\prime}\dot{\phi}_{\star},
\end{align}
 (with $B_{\star}^{\prime}=B^{\prime}(\phi_{\star})$) which integrates
to 
\begin{align}
\rho_{c\star}= & \rho_{c\star i}\left(\frac{r_{i}}{r}\right)^{3}e^{B_{\star}-B_{\star i}},
\end{align}
and therefore the ratio of the coupled collapsed matter to its background
(\ref{eq:rhoUbCb}-b) is also related to the radius through 
\begin{align}
\frac{\rho_{c\star}}{\rho_{c}}\propto & \left(\frac{a}{r}\right)^{3}e^{B_{\star}-B}.\label{eq:Cratio}
\end{align}

\subsubsection{Simplified evolution\label{sub:Simplified-evolution}}

Here we point out that this parallel evolution of coupled and uncoupled
collapse relies heavily on the implicit assumption that all species
collapse within the same radius, $r=r_{c}=r_{u}=r_{\phi}$. Introducing
the definition of the energy density contrast, 
\begin{align}
1+\delta_{m}= & \frac{\rho_{m\star}}{\rho_{m}}, & 1+\delta_{u}= & \frac{\rho_{u\star}}{\rho_{u}}, & 1+\delta_{c}=\frac{\rho_{c\star}}{\rho_{c}},\label{eq:overdensitiesDef}
\end{align}
 one can rewrite the radius of spherical collapse (\ref{eq:Uratio})
into 
\begin{align}
\left(\frac{a/a_{i}}{r/r_{i}}\right)^{3}(1+\delta_{i})= & 1+\delta_{u}.\label{rdefinition}
\end{align}
 Without loss of generality, the parallel between eqs. (\ref{eq:Uratio})
and (\ref{eq:Cratio}), and the total matter, can be expressed in
the relations 
\begin{align}
1+\delta_{c} & =(1+\delta_{u})\cdot e^{B_{\star}-B}, & 1+\delta_{m} & =(1+\delta_{u})\cdot\frac{\rho_{c0}\, e^{B_{\star}-B_{0}}+\rho_{u0}}{\rho_{c0}\, e^{B-B_{0}}+\rho_{u0}}.\label{deltac+t}
\end{align}
However these relations assume the initial conditions for $\delta_{c}$
and $\delta_{u}$ to be the same. We relax in Appendix \ref{sec:Unequal-initial-overdensities}
this condition, showing that the final results are not significantly
affected. For simplicity, we will keep assuming the equality of initial
conditions for the rest of the paper, as is implicit in previous results
\cite{Nunes:2004wn,Manera:2005ct,Tarrant:2011qe}.

We thus emphasize that the assumption, commonly made in the field
of DE spherical collapse \cite{Lokas:2000cn,Mainini:2002ry,Mainini:2003uf,Mota:2004pa,Lokas:2003cj,Solevi:2004tk,Solevi:2005ta,Nunes:2004wn,Manera:2005ct,LeDelliou:2005ig,Tarrant:2011qe},
setting the same radius of collapse for all species ($r=r_{c}=r_{u}=r_{\phi}$),
implies that all matter density contrasts are related to the uncoupled
case and their dynamics can be obtained through that of the uncoupled
matter. This simplifies greatly the treatment and governing equations
previously used for coupled and collapsing DE spherical collapse,
e.g. as in \cite{Nunes:2004wn,Manera:2005ct,Tarrant:2011qe}, since
one only needs to evolve the uncoupled DM overdensity equations to
get the coupled one via the above constraints. This means that we
can concentrate on the evolution of the uncoupled matter, and recover
the full evolution for all components.

\subsubsection{Other assumptions}

The central spherical patch is treated as a curvature varying Friedmann
model with the Raychaudhuri equation while the background is a regular
Friedmann model with lower density. The backreaction is neglected
invoking the Birkhoff theorem, assuming some extension of it to cosmological
backgrounds can be applied and the existence of a global separation
shell \cite{Mimoso:2009wj,LeDelliou:2011wk,LeDMM09b} within the boundary
between the two regions.

\subsubsection{Non-linear evolution}

We now will get the overdensity evolution equations for the uncoupled
DM component since the algebraic relation with the coupled overdensity,
from the unique radius constraint, renders the coupled evolution equations
superfluous.

Using eq. (\ref{rdefinition}), together with the Bianchi energy conservations
(eqs. \ref{eq:bkgU+C+DE}-a and \ref{eq:CollU}) for the Hubble parameters
of the background and collapsing regions yields the overdensity first
derivative 
\begin{eqnarray}
\dot{\delta}_{u} & = & 3(1+\delta_{u})\left[H-\mathcal{H}\right].\label{eq:deltaudot}
\end{eqnarray}
 The Einstein field equations in this frame are reduced to the Friedmann
and Raychaudhuri equations for the background scale factor $a$ (eqs.
\ref{eq:FriedmannRaychadhuri}) and the collapsing radius $r$: 
\begin{align}
\dot{\mathcal{H}}+\mathcal{H}^{2}= & -\frac{\kappa^{2}}{6}\sum\left(\rho+3P\right)_{\star},\label{eq:RayCollapse}
\end{align}
 to which the Klein-Gordon equations for the background (\ref{eq:KleinGordon})
and collapsing region are added in the form, in the case of totally
collapsing quintessence, 
\begin{align}
\ddot{\phi}_{\star}+3\mathcal{H}\dot{\phi}_{\star}+V_{\star}^{\prime} & =-\rho_{c\star}B_{\star}^{\prime}.\label{eq:KleinGordonCollapse}
\end{align}
 The case of the homogeneous quintessence obeys to the same equations
except that in eq. (\ref{eq:KleinGordonCollapse}),the collapsing
Hubble parameter is replaced by that of the background. The overdensity
evolution equation follows by differentiating eq. (\ref{eq:deltaudot})
\begin{eqnarray}
\left.\frac{\ddot{\delta}_{u}}{1+\delta_{u}}-\frac{\dot{\delta}_{u}^{2}}{(1+\delta_{u})^{2}}\right|_{u} & = & 3\left[\dot{H}-\dot{\mathcal{H}}\right].
\end{eqnarray}
Eventually one gets, together with using eqs. (\ref{deltac+t}-a,
\ref{eq:RayCollapse} and \ref{eq:FriedmannRaychadhuri}), (recall
from it and from eq. \ref{eq:rhoUbCb}-b that the coupled quantities
are linked with the uncoupled ones) 
\begin{align}
\ddot{\delta}_{u} & =-2\frac{\dot{a}}{a}\dot{\delta}_{u}+\frac{4}{3}\frac{\dot{\delta}_{u}^{2}}{1+\delta_{u}}+\frac{\kappa^{2}}{2}(1+\delta_{u})\left[\delta_{u}\rho_{u}+\delta_{c}\rho_{c}\right]+\frac{\kappa^{2}}{2}\left[\left(\rho_{\phi_{\star}}+3P_{\phi_{\star}}\right)-\left(\rho_{\phi}+3P_{\phi}\right)\right]\left(1+\delta_{u}\right).\label{eq:densityEvol}
\end{align}
If we have no coupled matter, that is if $\rho_{c}=0$, this is the
usual equation for the evolution the spherical collapse of uncoupled
matter as seen in \cite{Nunes:2004wn}. Following this evolution of
$\delta_{u}$ and using eq. (\ref{deltac+t}-b) we can re-obtain the
full matter evolution of \cite{Nunes:2004wn}.

\subsubsection{Linear evolution}

The linear evolution equations for the collapsing spherical patch
starts from the quintessence field linearisation with $\phi_{\star}=\phi+\delta\phi$
of the Klein-Gordon eq. (\ref{eq:KleinGordonCollapse}). We use the
linearisation of eqs. (\ref{deltac+t}) 
\begin{align}
\delta_{m,L} & =\delta_{u,L}+\frac{\Omega_{c0}}{\Omega_{c0}+\Omega_{u0}e^{B_{0}-B}}B^{\prime}\delta\phi, & \delta_{c,L} & =\delta_{u,L}+B^{\prime}\delta\phi,\label{eq:deltamdotL+cL}
\end{align}
  so we are left with the linear part of eq. (\ref{eq:KleinGordonCollapse})
\begin{gather}
\delta\ddot{\phi}+\left[3H+\rho_{c}B^{\prime}\right]\delta\dot{\phi}+\left[V^{\prime\prime}+\rho_{c}\dot{\phi}B^{\prime\prime}\right]\delta\phi+\dot{\delta}_{u,L}\dot{\phi}+\delta_{c,L}\dot{\phi}B^{\prime}\rho_{c}=0.\label{eq:KGlinear}
\end{gather}
 The linearisation of the field source term of eq. (\ref{eq:densityEvol})
yields 
\begin{align}
\left[\left(\rho_{\phi_{\star}}+3P_{\phi_{\star}}\right)-\left(\rho_{\phi}+3P_{\phi}\right)\right]_{L}= & 2\left(2\dot{\phi}\delta\dot{\phi}-V^{\prime}\delta\phi\right),
\end{align}
 so the density evolution linearises as (again recall that eq. \ref{eq:deltamdotL+cL}-b
relates coupled and uncoupled overdensities) 
\begin{align}
\ddot{\delta}_{u,L}= & -2\frac{\dot{a}}{a}\dot{\delta}_{u,L}+\kappa^{2}\left[2\dot{\phi}\delta\dot{\phi}-V^{\prime}\delta\phi\right]+\frac{\kappa^{2}}{2}\left[\delta_{u,L}\rho_{u}+\delta_{c,L}\rho_{c}\right].\label{eq:densityEvolL}
\end{align}
Again, if we have no coupled matter, this is the usual linear equation
for the spherical collapse evolution of uncoupled matter as seen in
\cite{Nunes:2004wn}. Following this evolution of $\delta_{u,L}$
and using eq. (\ref{eq:deltamdotL+cL}-a) we can re-obtain the full
matter linear evolution of \cite{Nunes:2004wn}.

\section{Numerical implementation and results: critical density evolution\label{sec:Numerical-implementation-and}}

In order to follow the evolutions of the segregated uncoupled matter/DM
spherical collapse for the four potentials we selected and inhomogeneous
coupled DM to DE models, a matlab code was produced ad hoc. Here we
present the results of our computations.

\subsection{Top hat and critical overdensity evolutions}

\subsubsection{Collapse of uncoupled matter/coupled DM}

\begin{figure}
\begin{centering}
\includegraphics[width=0.5\columnwidth]{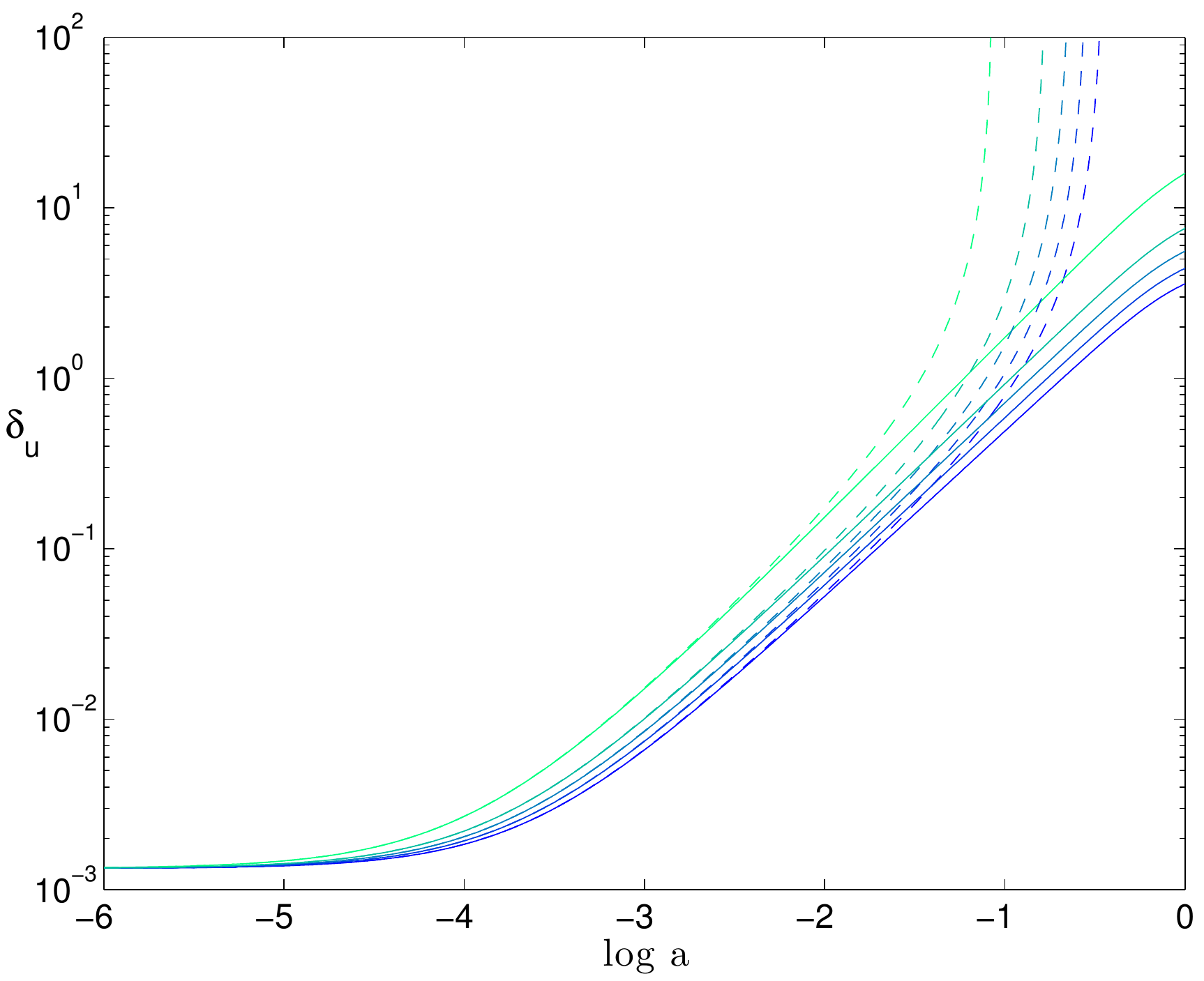} 
\par\end{centering}

\caption{\label{fig:Spherical-collapse-evolution}Spherical collapse evolution
of the uncoupled matter for the inhomogeneous double exponential potential.
We show both the linear (solid line) and the non-linear (dashed line)
evolutions with $\Omega_{u}=0.05$ and couplings $C=-0.2,-0.1,0,0.1$
and $0.2$ from bottom to top.}
\end{figure}

To provide the reader with comparison points to previous works \cite{Lokas:2000cn,Mainini:2002ry,Mainini:2003uf,Mota:2004pa,Lokas:2003cj,Solevi:2004tk,Solevi:2005ta,Nunes:2004wn,Manera:2005ct,LeDelliou:2005ig,Tarrant:2011qe},
we present here the diagram of one example of time evolution for the
linear and non-linear spherical top-hat collapse in a coupled inhomogeneous
quintessence model. We make the arbitrary choice of using the double
exponential potential, fixing the uncoupled matter to baryons, $\Omega_{u}=0.05$,
for a range of coupling $C\in\left[-0.2;0.2\right]$. We display this
plot of $\delta_{u}$ vs $\log a$ in figure  \ref{fig:Spherical-collapse-evolution}
. It shows that the linear values of overdensities go up with coupling,
and consequently that the corresponding collapse time decreases, as
expected \cite{Nunes:2004wn,Manera:2005ct,Tarrant:2011qe}. This appears
mainly as a consequence of background evolution, as induced from eq.
(\ref{eq:rhoUbCb}-b). The effect of inhomogeneous quintessence also
speeds up collapse time, as seen in the last terms of eq. (\ref{eq:densityEvol})
or the second linear terms of eq. (\ref{eq:densityEvolL}). This behaviour
is similar for the other potentials and varying $\Omega_{u}$ with
a fixed $C$ gives the same or opposite behaviour depending on the
sign of the chosen $C$ (for more details on this $\Omega_{u}-C$
degeneracy , see section   \ref{sub:critical-overdensities-evolution}).

Note that the effect of the coupling parameter $C$ on the evolution
of the coupled matter energy density comes through term $e^{-C\kappa\Delta\phi}$,
as seen in eq.~(\ref{eq:rhoUbCb}), so that the effect comes through
a combination of the parameter $C$ and the range of the scalar field
$\phi$ evolution. This means that the same value of the coupling
parameter $C$ can have different effects depending on the type of
scalar potential used. For the four potentials we use, the range of
the scalar field between matter-radiation equality and the present
is approximately $\Delta\phi=1.6$, $\Delta\phi=1.8$, $\Delta\phi=2.7$
and $\Delta\phi=1.5$ for the two exponentials, Albrecht-Skordis,
SUGRA and Steinhardt potentials respectively. We thus expect that
the coupling strength is roughly equivalent between our four potentials,
though slightly enhanced for the SUGRA potential. Note that the value
of $\Delta\phi$ will also depend on the amount of coupling used,
the quotes are for the mean of the evolutions we used numerically.

\subsubsection{Relative linear overdensities evolution}

\begin{figure}
\begin{centering}
\includegraphics[width=0.49\columnwidth]{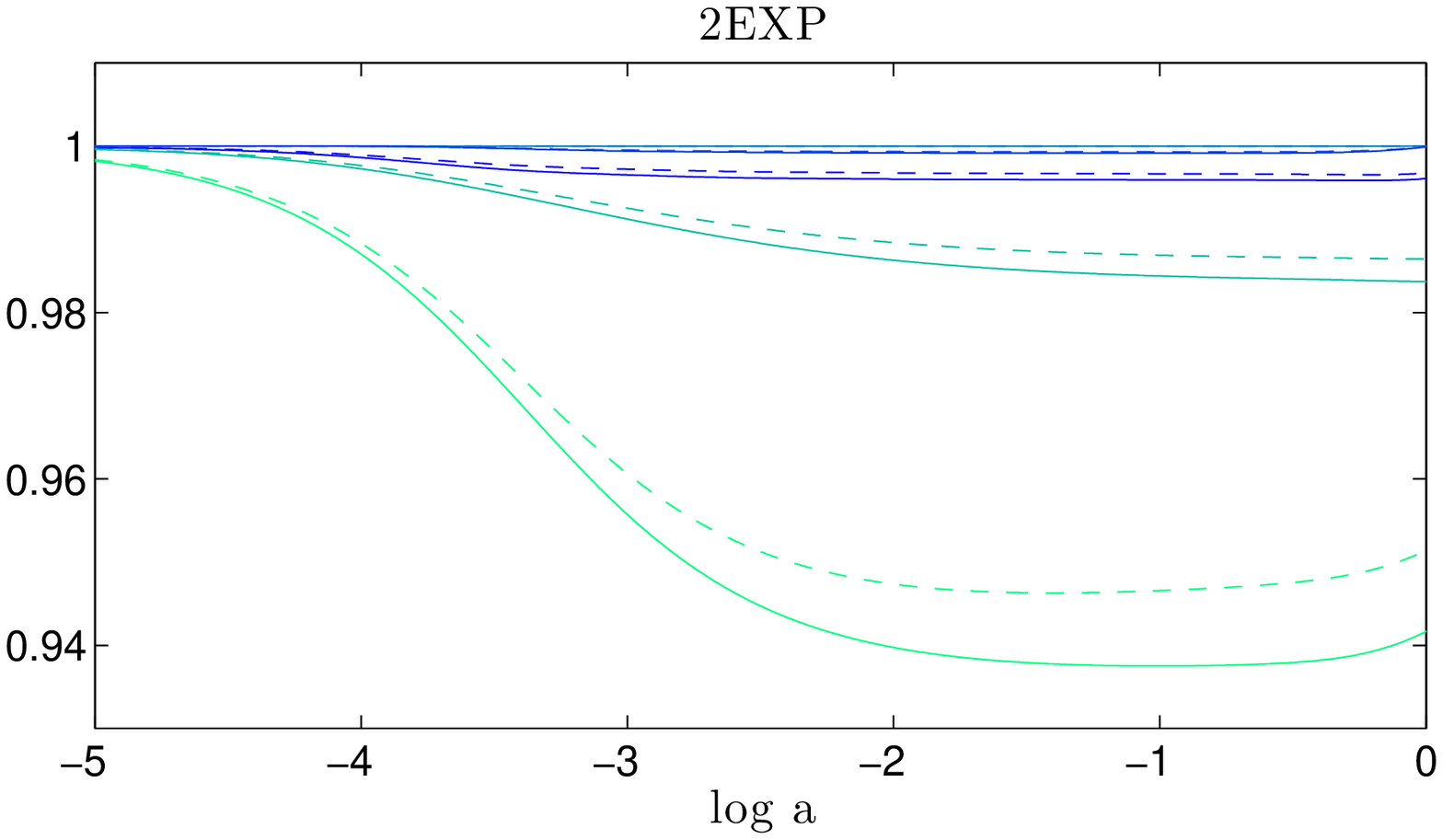}\includegraphics[width=0.49\columnwidth]{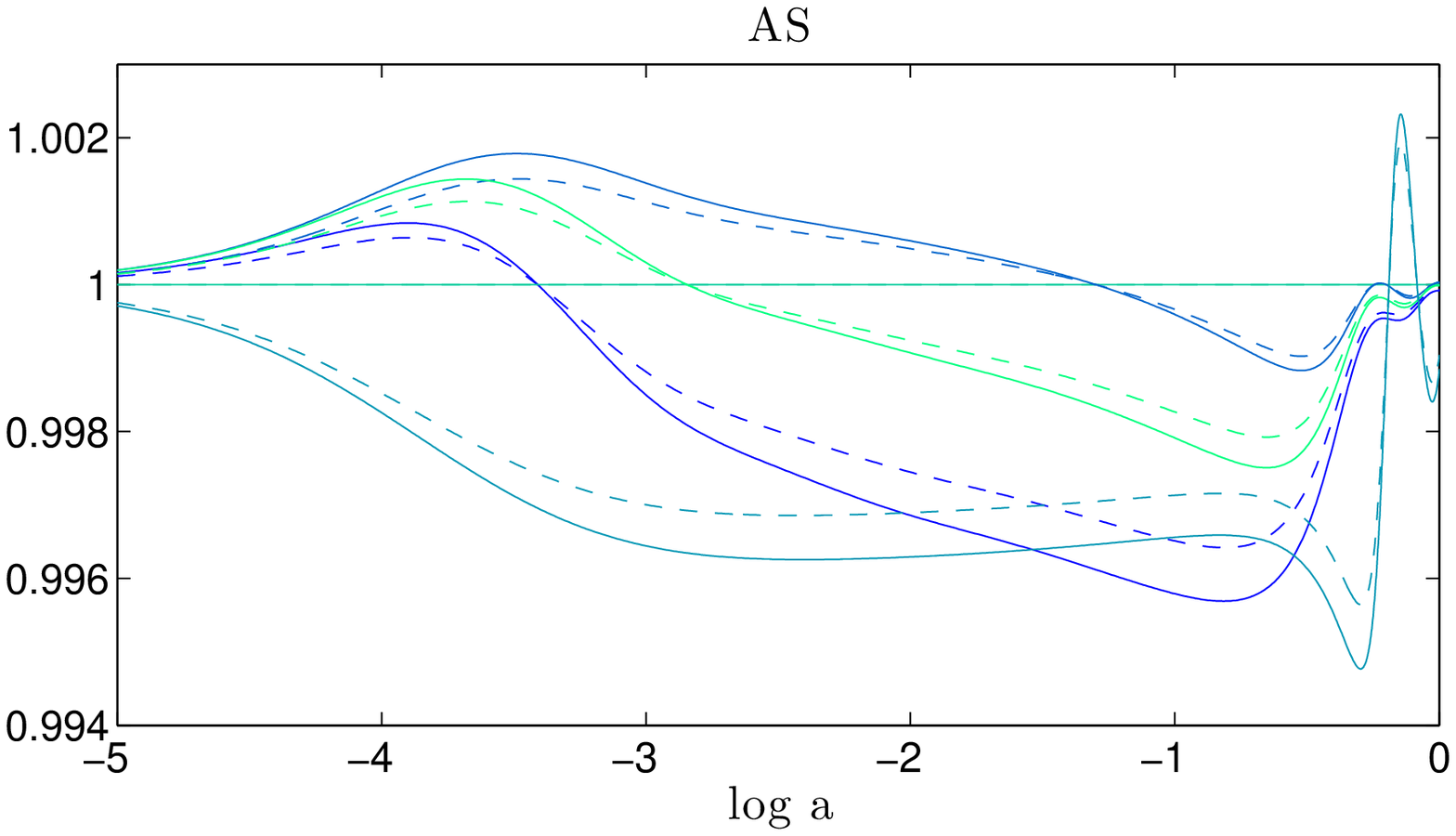}\\
 \includegraphics[width=0.49\columnwidth]{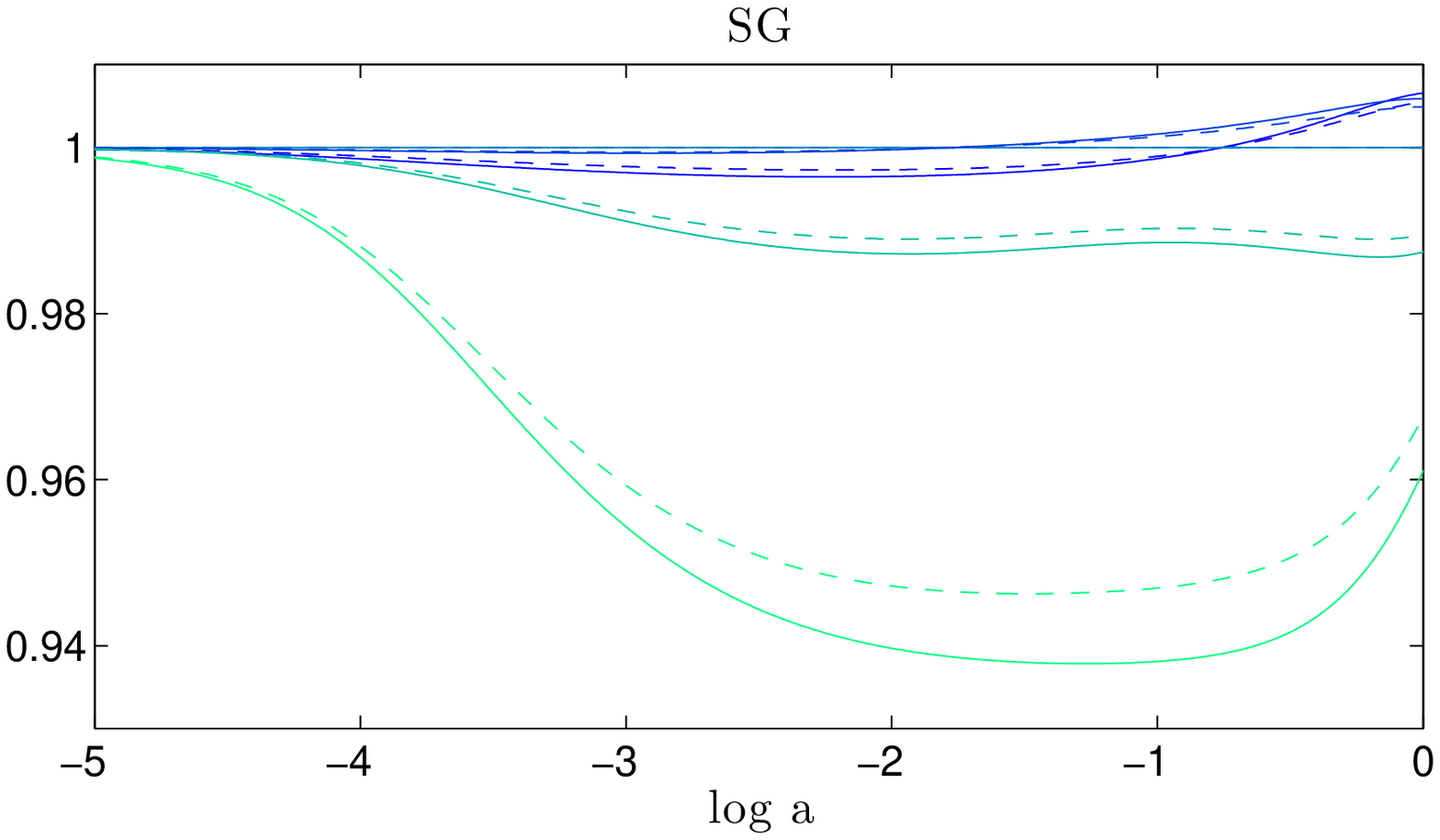}\includegraphics[width=0.49\columnwidth]{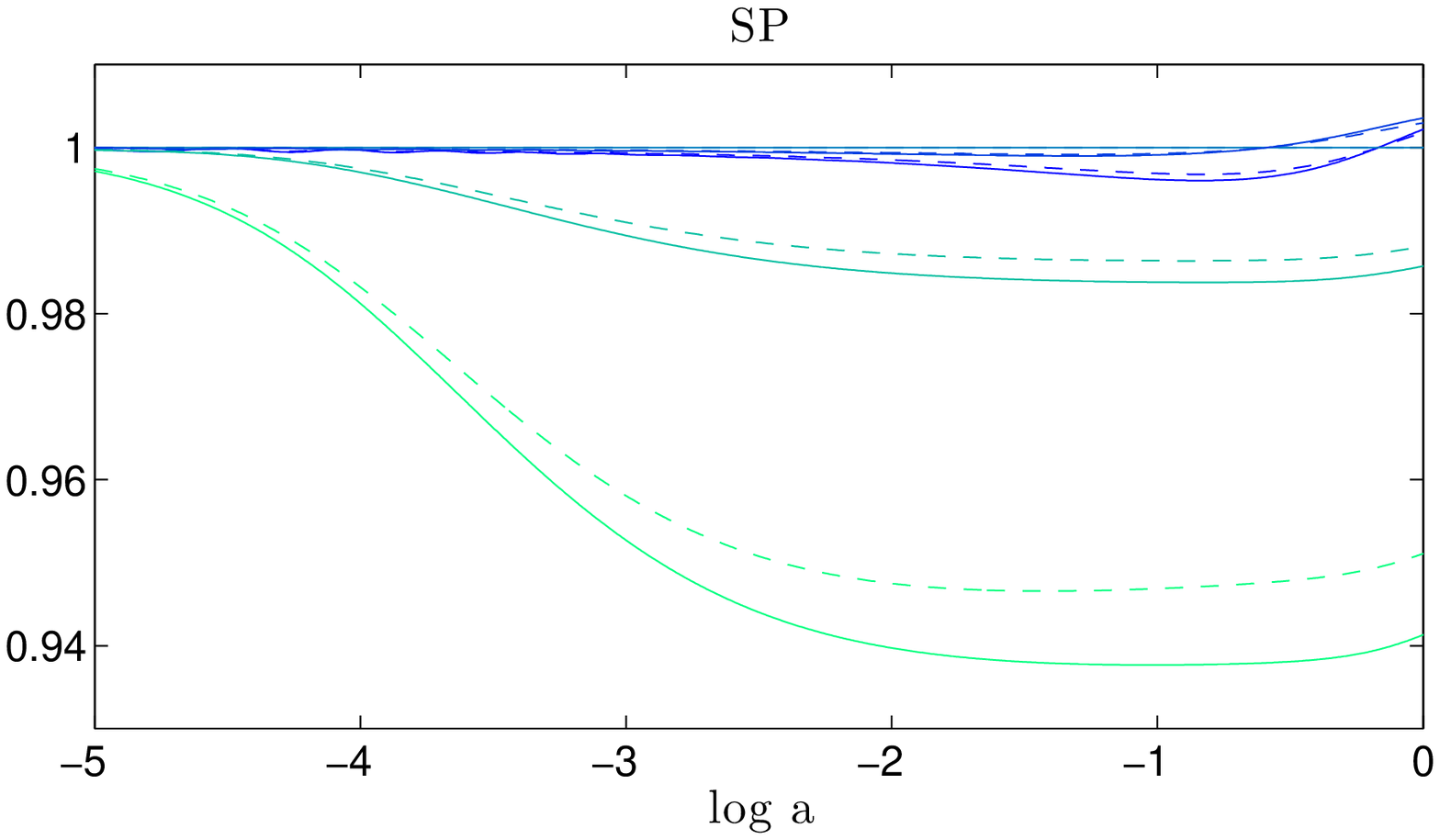} 
\par\end{centering}

\caption{\label{fig:Relative-linear-overdensity}Relative linear overdensity
evolution for $\delta_{c\star,L}/\delta_{u\star,L}$ (solid line)
and $\delta_{m\star,L}/\delta_{u\star,L}$ (dashed line), 
for the potentials: double exponential (2EXP), Steinhardt (SP), SUGRA
(SG), Albrecht-Skordis (AS). The values of coupling are $C=-0.2,-0.1,0,0.1$
and $0.2$ (dark to light) for all the potentials except Albrecht-Skordis
where $C=-0.2,-0.1,0$ and $0.05$.}
\end{figure}

As we have followed separately the coupled (DM) and uncoupled (including
baryons) matter components, as opposed to previous works \cite{Nunes:2004wn,Manera:2005ct,Tarrant:2011qe},
it is interesting to confront all the models used in their relative
linear overdensity evolution with respect to the uncoupled component.
Note that although we are able to reproduce the results from \cite{Nunes:2004wn,Manera:2005ct,Tarrant:2011qe},
(a) we do so in a much simpler way as described in section \ref{sub:Simplified-evolution},
and (b) we focus on the coupled/uncoupled matter differences which
are not discussed in those previous works. In figure  \ref{fig:Relative-linear-overdensity},
we display the evolutions of the coupled and total overdensities relative
to the uncoupled one, for the models involving the four potentials
described in section   \ref{sub:Choice-of-potential} in the tracking
regime, and for a range of coupling. Again, we fix the uncoupled matter
to baryons, $\Omega_{u}=0.05$, for a range of coupling $C\in\left[-0.2;0.2\right]$
for each model.

In all cases, absence of coupling makes all species behave the same
way (we see, as expected, an horizontal line). Because the total overdensity
combines coupled and uncoupled ones, as from eqs. (\ref{eq:overdensitiesDef})
\begin{align}
\frac{\delta_{m}}{\delta_{u}}= & \frac{\delta_{c}}{\delta_{u}}+\frac{\rho_{u}}{\rho_{m}}\left(1-\frac{\delta_{c}}{\delta_{u}}\right),\label{eq:RelativeOverdensities}
\end{align}
we see that its departure from $\delta_{u}$ is always less than that
of $\delta_{c}$.  This departure is small when varying $C$, however
in the range of chosen variations, it appears markedly smaller for
$\delta_{m}$ than for $\delta_{c}$. In other words, for $C>0$ we
have $\delta_{c}<\delta_{m}<\delta_{u}$, while for $C<0$ we have
$\delta_{c}>\delta_{m}>\delta_{u}$. Negative couplings, with the
form of eq. (\ref{eq:bkgU+C+DE}-b) and $B^{\prime}=-C\kappa$, seem
to induce stronger collapse in the coupled DM than in the uncoupled
matter, whereas positive couplings have the opposite effect, although
with a much stronger discrepancy than for the negative couplings in
all models. This can also be interpreted from eq. (\ref{deltac+t}-a)
as $\phi_{\star}-\phi>0$. The qualitative behaviour for the double
exponential, Steinhardt and SUGRA potentials is quite regular. However
the Albrecht-Skordis potential displays quite strong oscillations
in the linear overdensity near late times, reflecting its nature driving
the field towards its local minimum around which it oscillates. This
will be discussed further in section   \ref{sub:critical-overdensities-dependenc}.

\subsubsection{Critical overdensities evolution\label{sub:critical-overdensities-evolution}}

We now present the diagrams of critical overdensities as a function
of their collapse redshift for the four potentials. This time again,
we fix the uncoupled matter to baryons, $\Omega_{u}=0.05$, for a
range of coupling $C\in\left[-0.2;0.2\right]$ for each model.

Note that it can be shown that fixing the coupling respectively to
a positive or negative value, for a range of uncoupled matter parameter
$\Omega_{u}\in\left[0.05;0.3\right]$ can mimic the effect of varying
the coupling. The interval is chosen such that the lower limit restricts
uncoupled matter to measured baryons, while the upper limit corresponds
to no coupled matter (recall that we fix $\Omega_{m}=\Omega_{u}+\Omega_{c}=0.3$).
With such parameterisation, growing $\Omega_{u}$ is indirectly equivalent
to decreasing the amount of coupling $\left|C\right|$. Thus varying
$\Omega_{u}$ can be mapped to varying $C$, accounting for the sign
of the constant $C$ chosen. This behaviour seems to extend the dark
degeneracy found at the homogeneous level \cite{Kunz:2007rk}.  However,
the spatial segregation between the coupled and uncoupled matter components
illustrated in the overdensities opens the door for mass determinations
of each component separately, if the segregation is clear enough.
Moreover, studies such as \cite{Bertolami:2007zm,Bertolami:2007tq,Bertolami:2012yp,Abdalla:2007rd,Abdalla:2009mt}
access directly the value of the coupling through the virial ratio
of clusters, so that degeneracy should, in principle, be broken.

\begin{figure*}
\includegraphics[width=0.5\columnwidth]{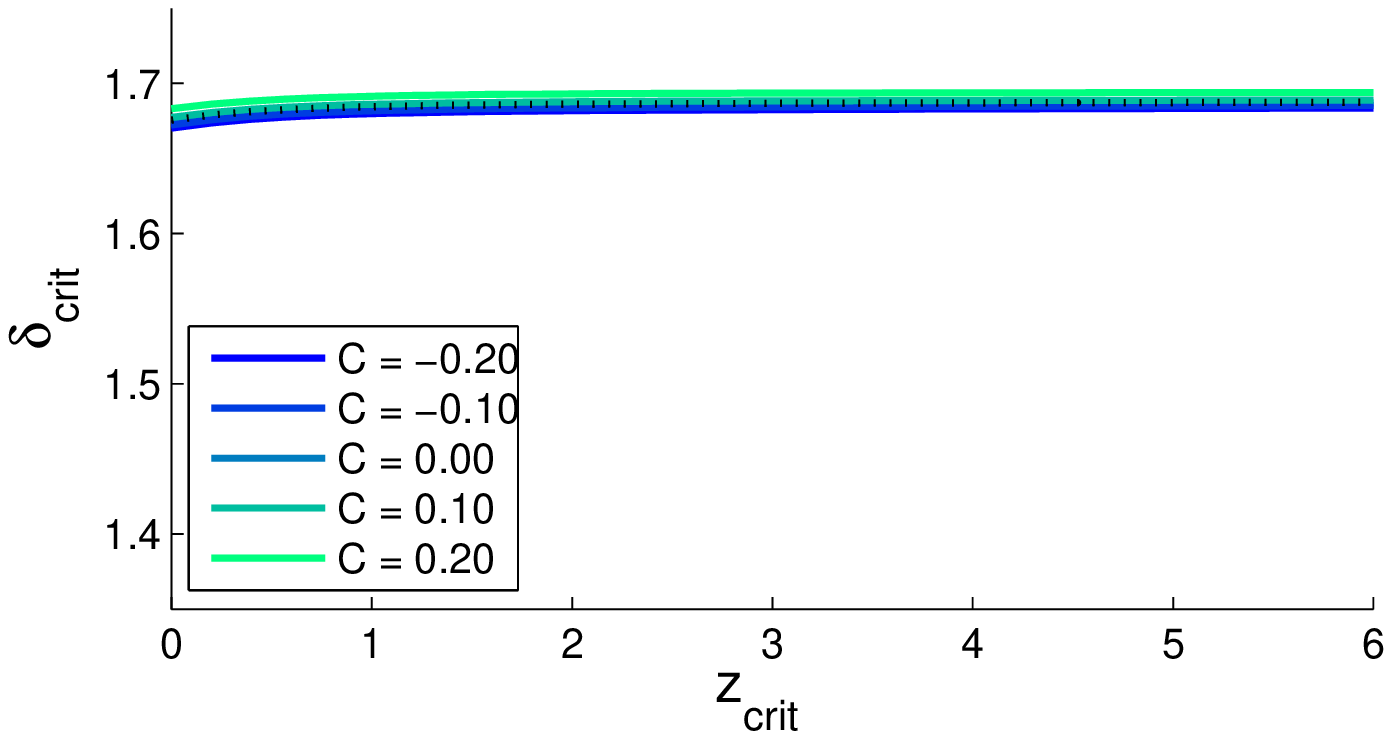}\includegraphics[width=0.5\columnwidth]{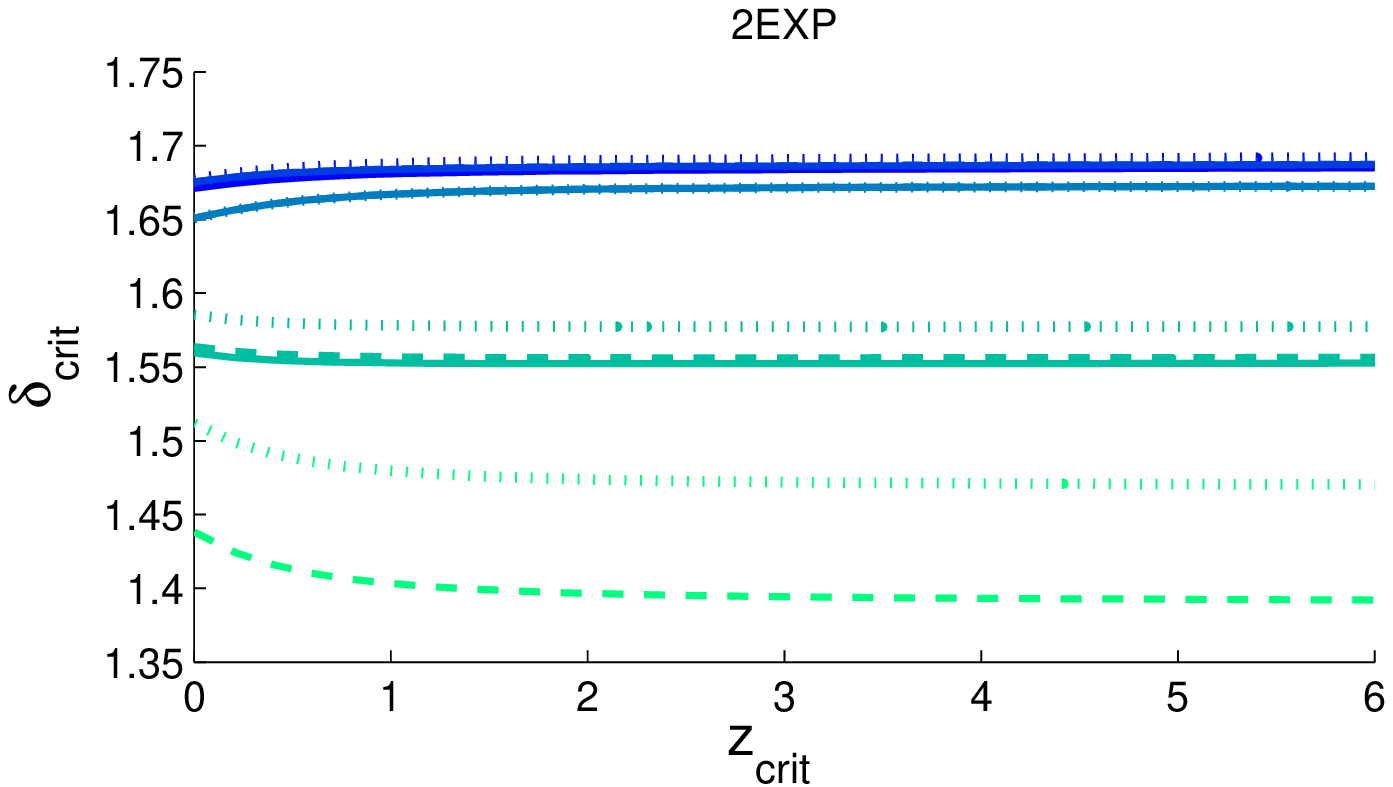}\\
 \includegraphics[width=0.5\columnwidth]{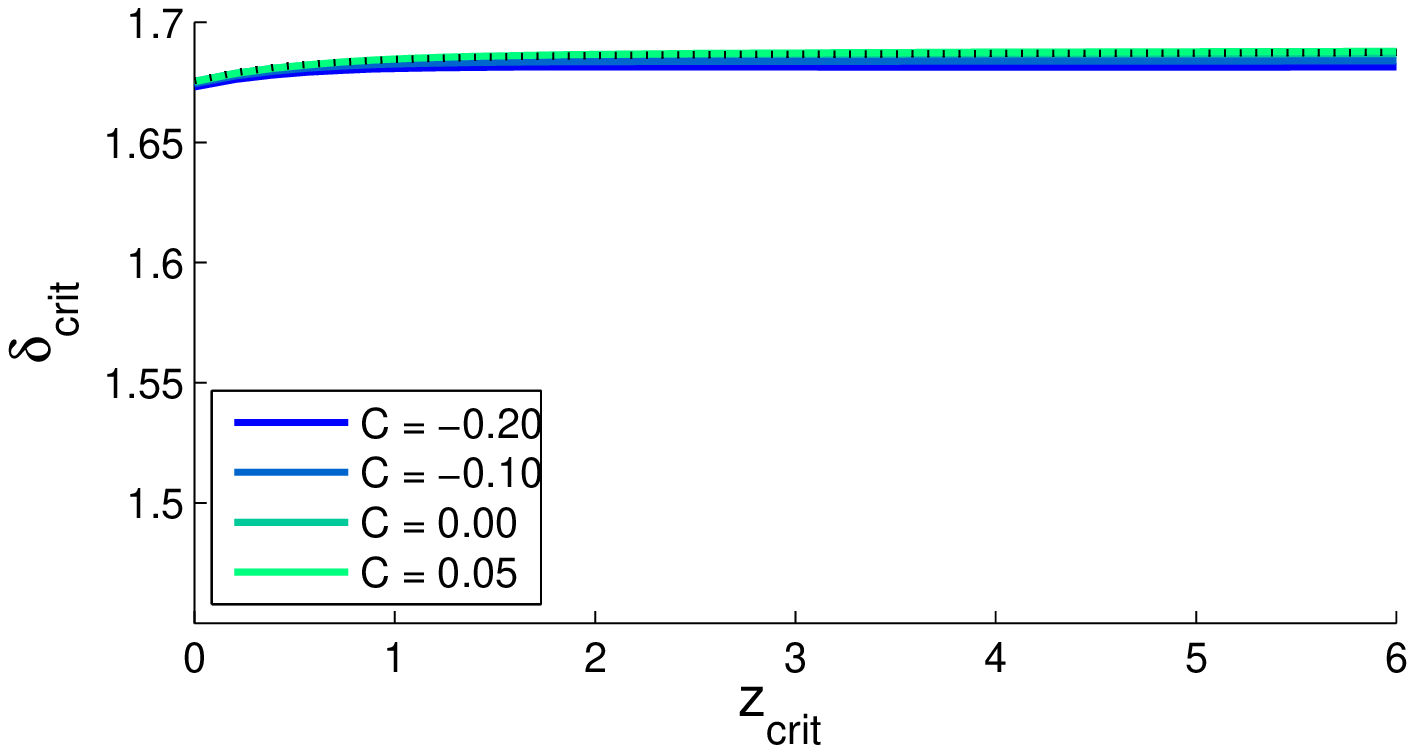}\includegraphics[width=0.5\columnwidth]{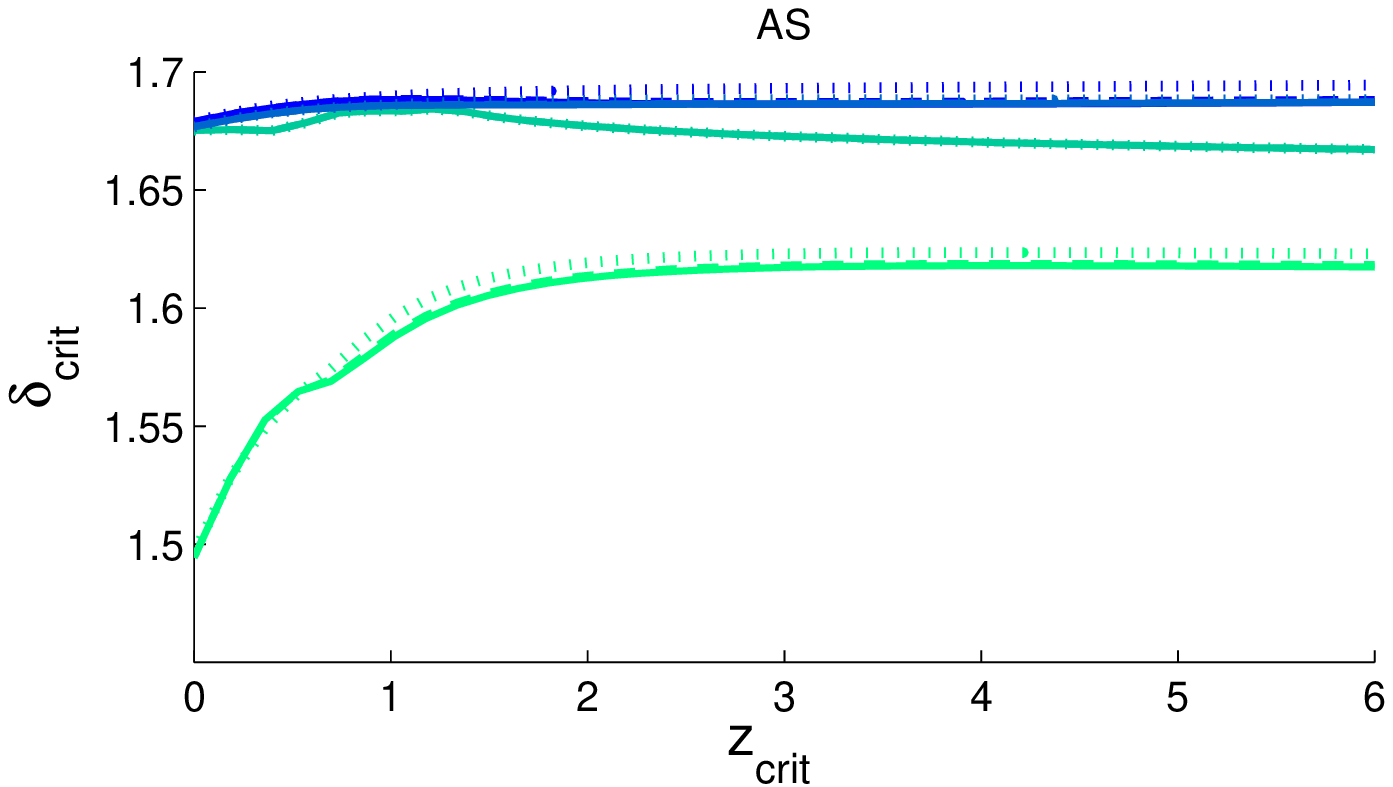}\\
 \includegraphics[width=0.5\columnwidth]{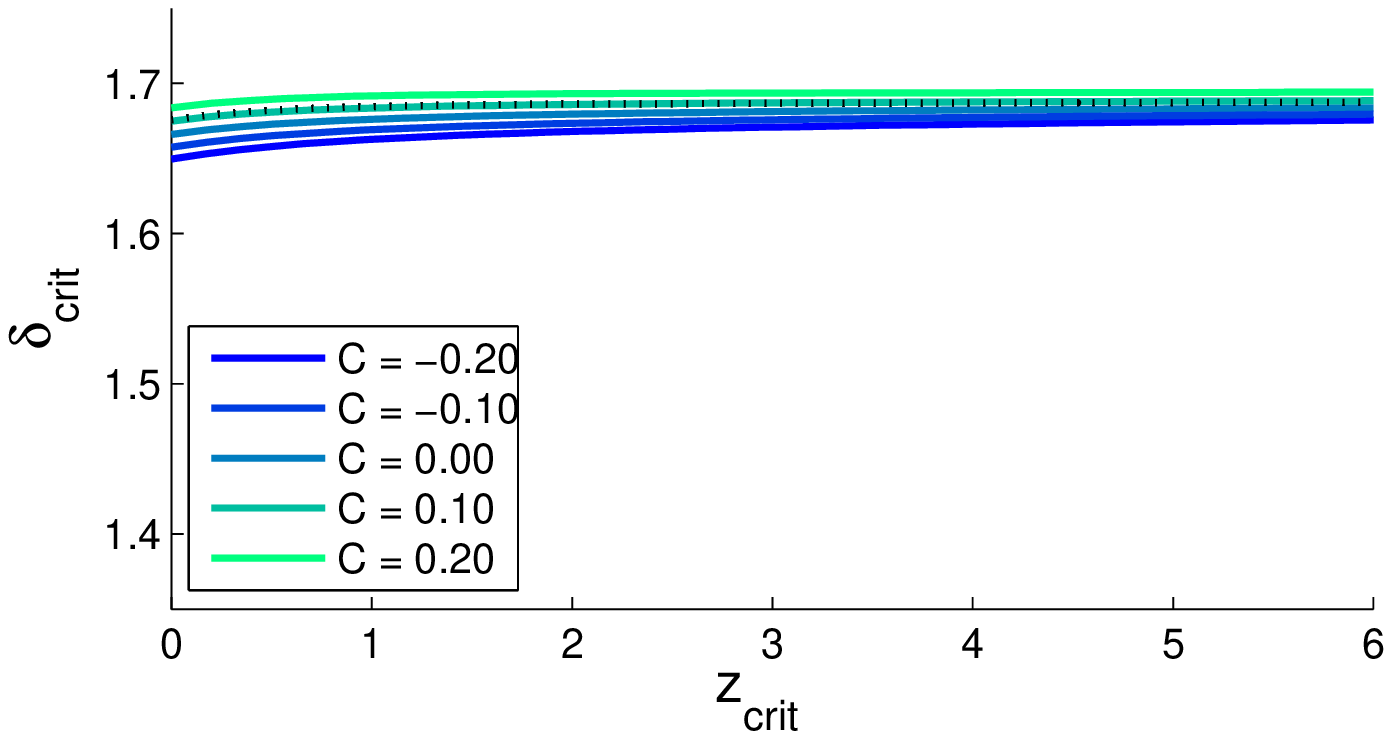}\includegraphics[width=0.5\columnwidth]{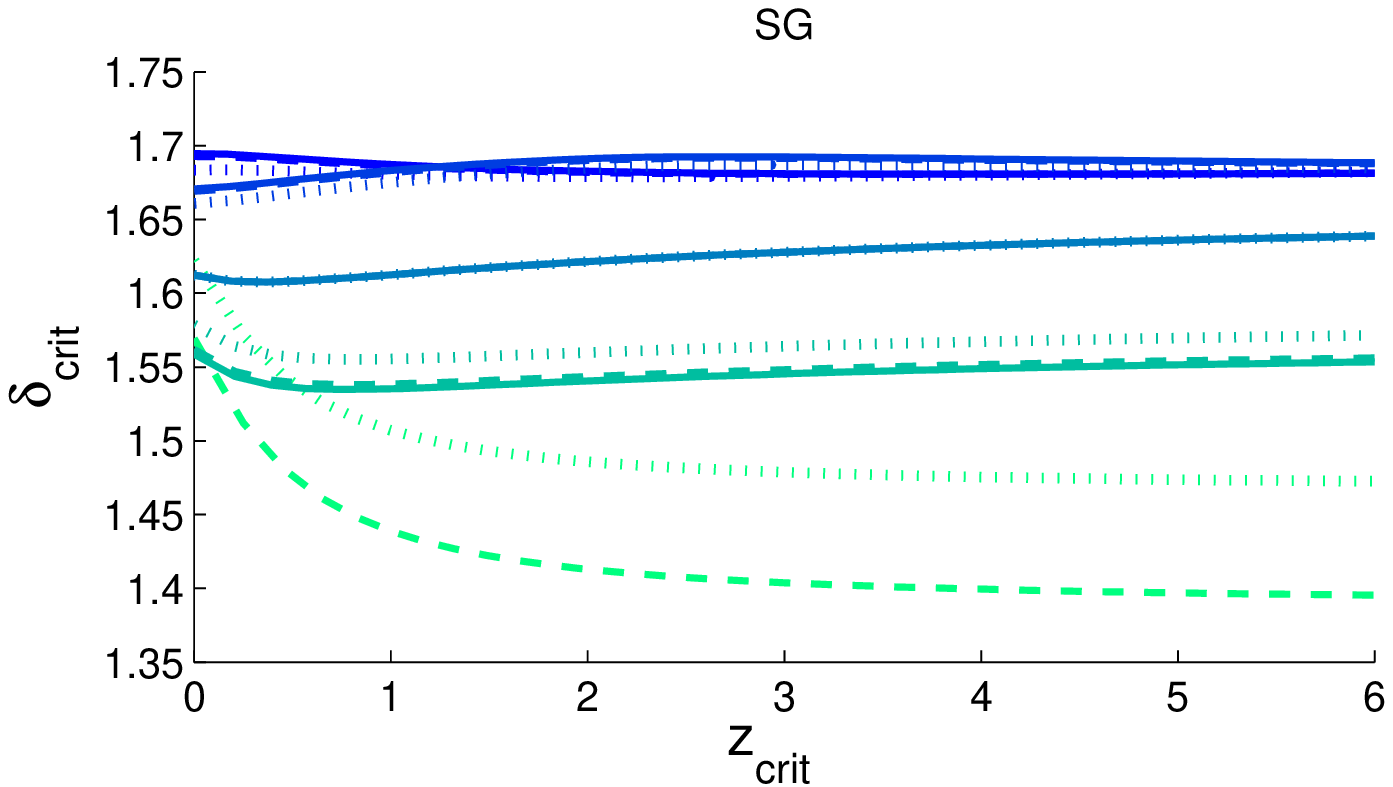}\\
 \includegraphics[width=0.5\columnwidth]{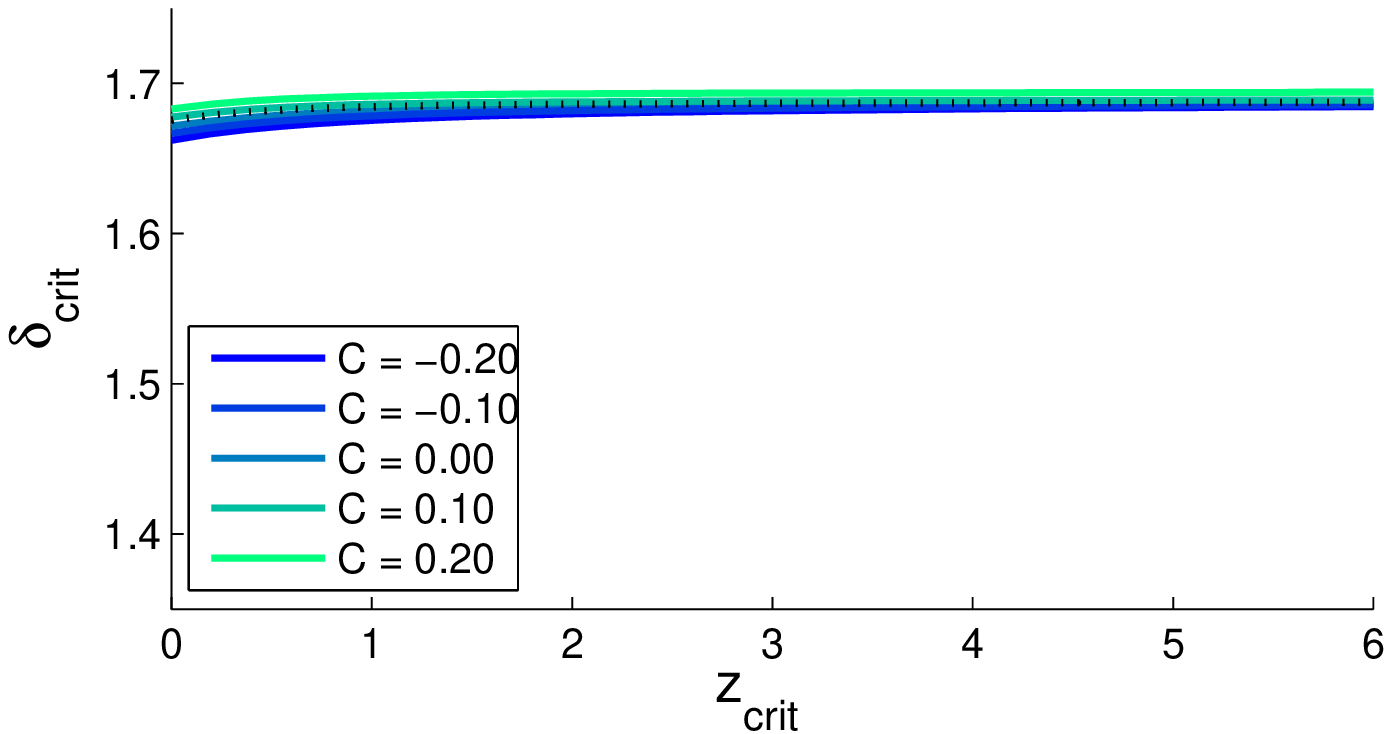}\includegraphics[width=0.5\columnwidth]{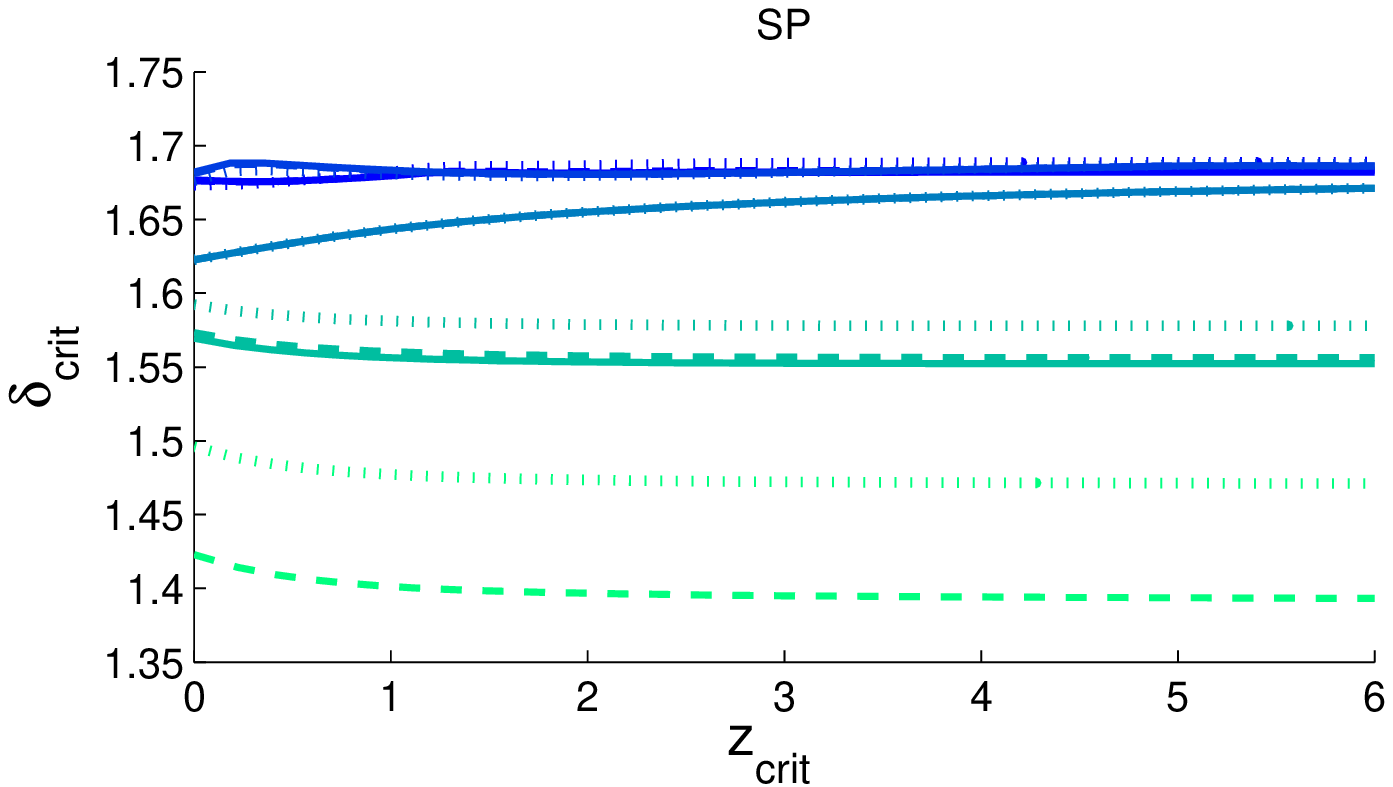}

\caption{\label{fig:Critical-overdensities-of}Critical overdensities as a
function of the non-linear collapse redshift for the double exponential
(2EXP), Albrecht-Skordis (AS), SUGRA (SG) and Steinhardt (SP) potentials.
The left panel represents a homogeneous quintessence model with the
total matter density contrast (solid lines) and as a reference the
$\Lambda$CDM result (dotted line). The right panel represents an
inhomogeneous quintessence model, with the density contrasts for uncoupled
matter ( dotted line), coupled matter (solid line) and total matter
(dashed line). The curves lightness reflect the values of $C$.}
\end{figure*}

In figure  \ref{fig:Critical-overdensities-of} 
, the leftmost panels give the homogeneous quintessence models behaviour
while the rightmost panels recall the total matter (dashed line) collapse
of the inhomogeneous coupled model for comparison with previous works
\cite{Mota:2004pa,Nunes:2004wn,Manera:2005ct,Tarrant:2011qe}. These
panels also provide the uncoupled (dotted) and coupled (solid) evolutions.
The scale and range of the diagrams is kept for each panel within
a row for comparison purpose.

As is now known \cite{Nunes:2004wn,Manera:2005ct,Tarrant:2011qe},
the variations induced by turning on coupling in homogeneous quintessence
are markedly smaller than that induced in collapsing models (compare
left and right, dashed lines, panels). Similarly as seen in figure
 \ref{fig:Spherical-collapse-evolution}, increasing the value of
$C$ both increase the corresponding value of critical linear overdensity
and the collapse redshift.%

Thanks to the simplifying assumption that leads to eqs. (\ref{deltac+t}),
the dynamics needs only to be followed in $\delta_{u,crit.}$, shown
in the right panel. It displays, in the negative coupling range, a
lower collapse threshold $\delta_{u,crit.}$ for uncoupled matter
than for coupled DM, consistent with eq. (\ref{eq:RelativeOverdensities}),
with a sharper decrease towards later times for uncoupled matter.
The picture is inverted in the positive coupling range: the DM collapse
threshold $\delta_{c,crit.}$ is lower than for uncoupled matter  in
that case and the increase towards later times is sharper than for
uncoupled matter. This vindicates the claim for baryon-DM segregation
with coupled DE.%

\subsection{Critical overdensities dependence on coupling\label{sub:critical-overdensities-dependenc}}

\subsubsection{General remarks}

\begin{figure*}
\noindent \begin{centering}
\hspace{-1cm}\includegraphics[width=0.36\textwidth]{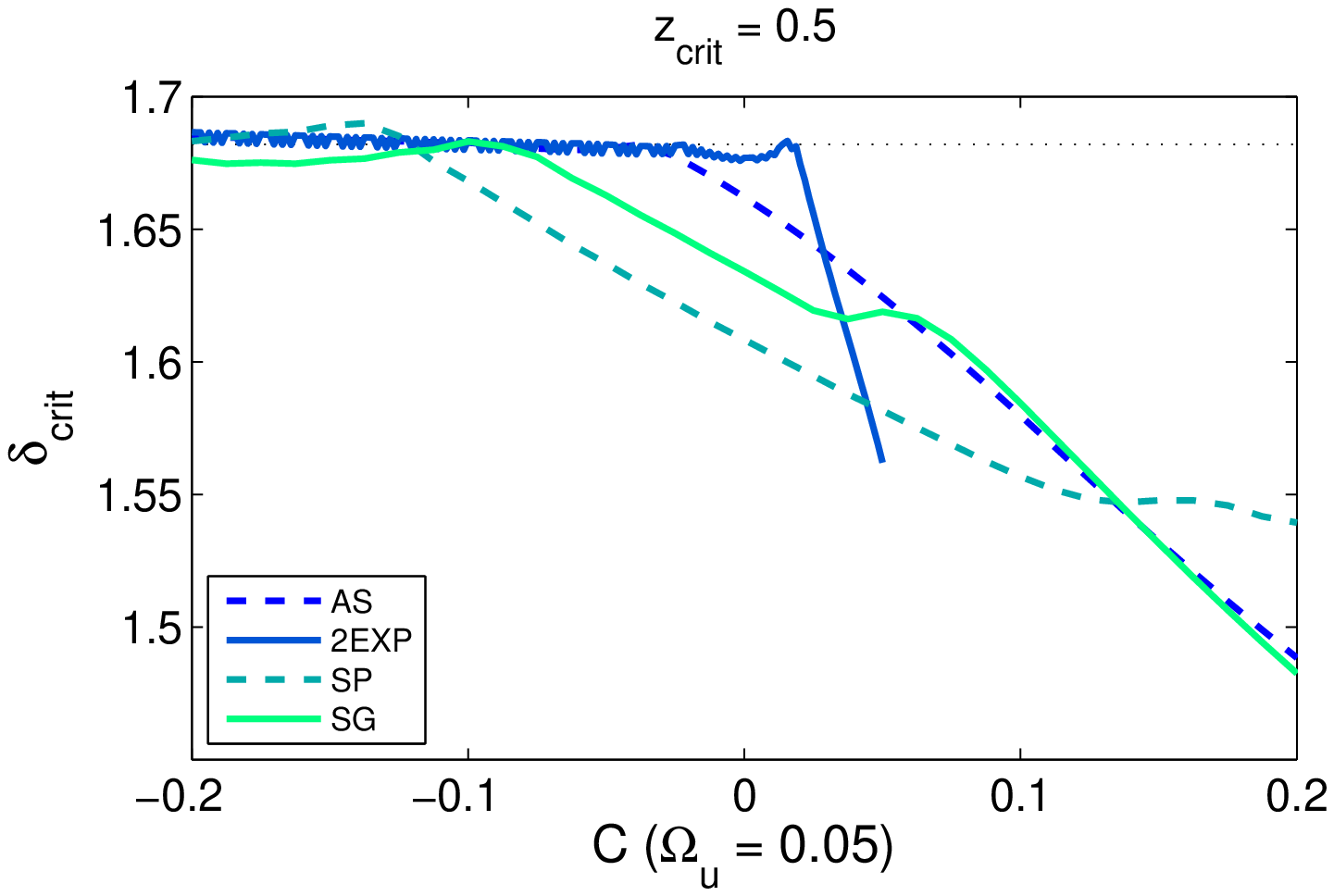}\includegraphics[width=0.36\textwidth]{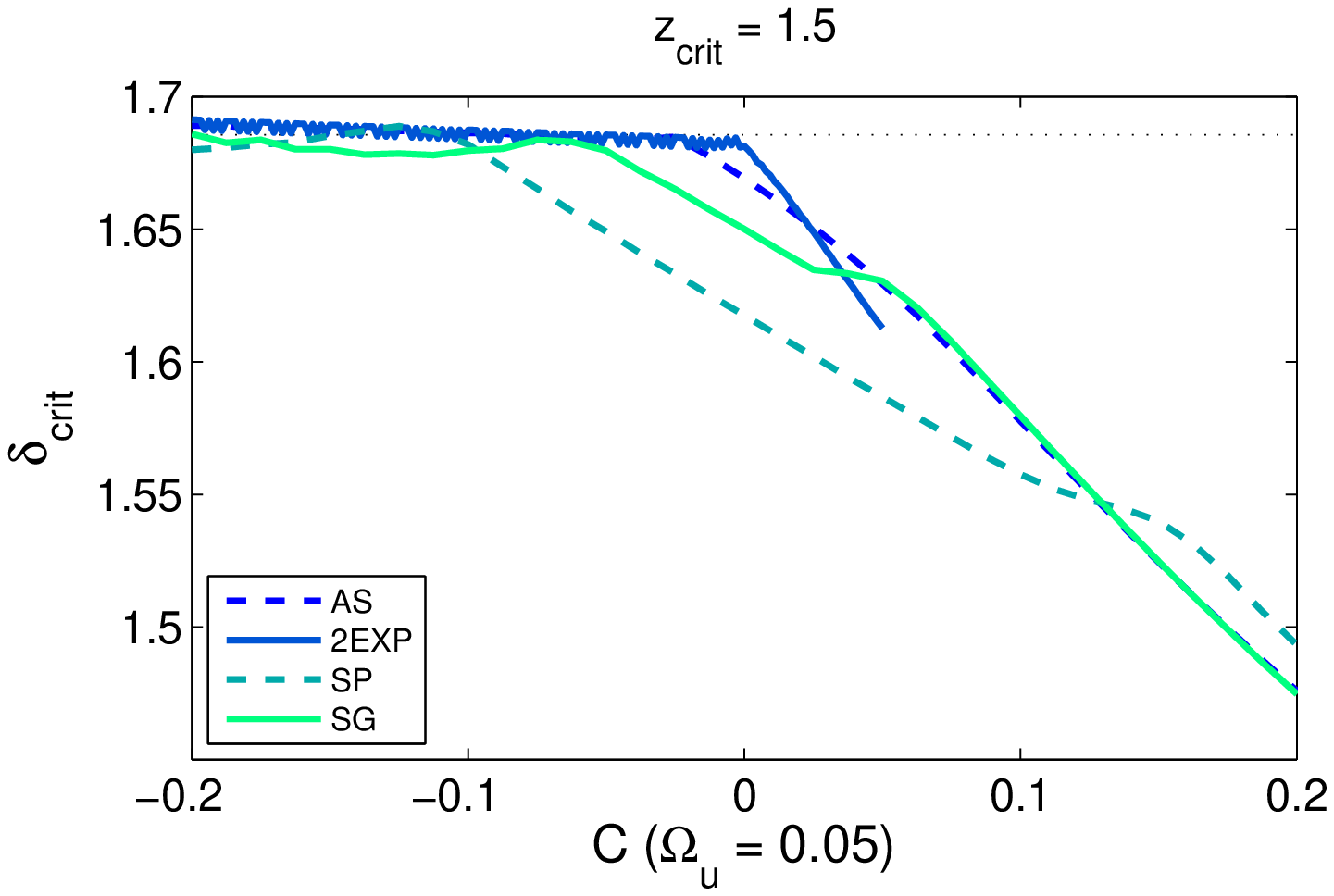}\includegraphics[width=0.36\textwidth]{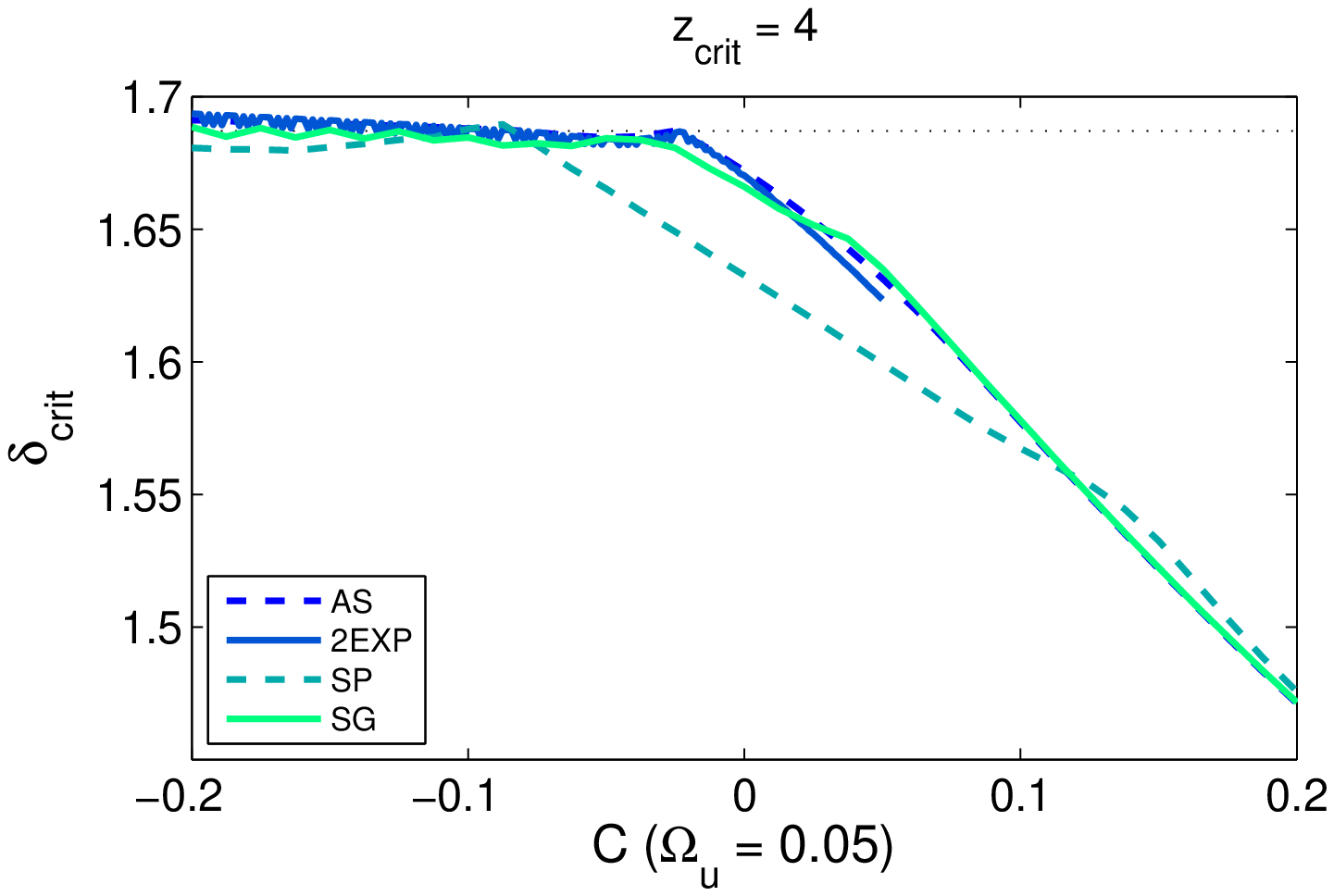}
\par\end{centering}

\caption{\label{fig:Critical-overdensities-synthesis} Value of the uncoupled
matter critical overdensity  as a function of the coupling at a reference
critical redshift of $z=0.5$, $z=1.5$ and $z=4$. We use $\Omega_{u}=0.05$.
Each panel shows the results for the four potentials we study, from
bottom to top: SUGRA, Steinhardt, double exponential and Albrecht-Skordis.
The dotted line shows the critical overdensity value at the reference
redshift in a $\Lambda$CDM model.}
\end{figure*}

We synthesise our results for all the potentials in figure  \ref{fig:Critical-overdensities-synthesis},
representing their variations in critical overdensities as a function
of coupling.

Note that the features of each potential in figure  \ref{fig:Critical-overdensities-synthesis}
agree with those of figure  \ref{fig:Critical-overdensities-of}.
Those features seems to characterise each potential.

A striking first observation is that, at fixed time ($z_{crit.}$),
the critical overdensities generally are decreasing with $C$, in
agreement with section   \ref{sub:critical-overdensities-evolution}'s
considerations. This can be understood as fixing $z_{crit.}$ requires
to vary initial $\delta_{i}$ to maintain that collapse time so as
to compensate the effect of the variations of $C$ in the exact opposite
way. From figure  \ref{fig:Spherical-collapse-evolution}, increased
$C$ leads to decreased collapse time, so a smaller initial $\delta_{i}$
is required to compensate that effect, leading to a smaller final
linear overdensity.

Recall from section \ref{sub:critical-overdensities-evolution} that
increasing $\Omega_{u}$ decreases the amount of coupling $\left|C\right|$.
In particular, the curve features found for SP, AS, 2EXP and SG in
the respective $C>0$ and $C<0$ ranges of the panels can be mapped
in corresponding varying $\Omega_{u}$ diagrams.

\subsubsection{discriminating models}

For larger values of z all potential behave similarly while at smaller
redshifts, each potential displays individual features due to the
emergence of DE dominated behaviour.

The 2EXP potential displays sections, mostly in the $C>0$ part, where
it also behaves linearly, with a larger slope than the SG. This can
be conjectured as reflecting the scaling solution behaviour. However
it has an hybrid behaviour as its critical overdensity asymptotes
to a constant (the $\Lambda$CDM critical overdensity) in the $C<0$
region. That behaviour can be attributed to oscillations around its
dominating solution.

The SP and SG potentials offer the most features: they match the 2EXP
linear $C>0$ behaviour as well as its more or less constant $\Lambda$CDM
asymptote $C<0$ region at large $\left|C\right|$ but shows an intermediate
region with similar slope. One can conjecture a link between them,
both potentials being built from inverse powers: the SG, with an exponential
cutoff, the SP, of an infinite sum of inverse powers.

Finally, the AS potential brings the most untypical features: it doesn't
admit solutions for $C>0.05$ and it decreases faster than any other
model for $C>0$. Whereas for $C<0$ it oscillates and asymptotes
to a constant $\Lambda$CDM value, like the 2EXP potential. This can
be interpreted from the fact that the AS potential offers a local
minimum for the field to be trapped in and oscillate around. From
eqs. (\ref{eq:bkgU+C+DE}-b,c), we see that $C>0$ feeds more energy
in the DE sector (assuming DE slow roll so $\dot{\phi}>0$ as all
chosen potentials decay with field, in a first approximation), thus
feeding the oscillations, whereas $C<0$ has the opposite energy transfer
and thus dampens the oscillations around the potential local minimum,
where the model effectively behaves like a cosmological constant.
Thus similarly as with 2EXP but in a stronger way, as the oscillation
behaviour is stronger in AS than 2EXP, we have the dampened behaviour
in the constant asymptote region and the excited behaviour in the
negative slope region.

Of course this requires deeper investigations.

\section{Discussion and conclusion\label{sec:Discussion-and-conclusion}}

In this work we have presented results supporting the segregation
of uncoupled matter and coupled DM in interacting DE models and have
done so with the exposition of an innovative and simplifying approach
to the dynamics of such system. The method employed relies on the
assumption common in the field that all species collapse within the
same top hat radius. The novel treatment comes from relying on the
algebraic relations between species overdensities derived from that
assumption to reduce the need to follow complex dynamics of the collapse
to the simpler evolution of uncoupled matter. This also clarifies
the reason why most of the effects we found are mainly driven by background
behaviours, principal components in the evolution equation outside
of the uncoupled overdensity. Exploring the segregated spherical collapse,
we made contact with previous work and established, in this model,
the difference in behaviour between uncoupled matter and DM interacting
with DE from (a) the differences in linear overdensity evolution between
species, (b) the differences in critical overdensity functions between
species, and (c) an original synthesis of critical overdensity as
a function of coupling, directly or indirectly. On a smaller note,
we also established the equivalence between variations of the coupling
and variations of the amount of uncoupled DM. This degeneracy is reminiscent
of \cite{Kunz:2007rk} and possibly arising from the constraint of
equal radii for all species. We expect it to be broken by (a) independent
mass determination of the uncoupled and coupled components with clear
enough segregation as argued in \cite{Peebles:2009th}, and (b) direct
coupling evaluation as in virial cluster studies \cite{Bertolami:2007zm,Bertolami:2007tq,Bertolami:2012yp,Abdalla:2007rd,Abdalla:2009mt}.
In the interaction model we chose, positive coupling corresponds to
feeding the DE sector from the coupled DM density whereas negative
coupling induces the opposite. This is a cornerstone to understand
(a) that negatively coupled DM collapses more than uncoupled matter
while the opposite is true of positively coupled DM; (b) that instabilities
in the behaviour of some of the chosen potentials are enhanced by
positively coupled DM collapse while the negatively coupled DM dampens
them. This confirms the potential detectability of dark sector coupling
in a similar fashion as from the tidal streams of stars in the Sagittarius
dwarf galaxy due to DM segregation \foreignlanguage{british}{\cite{Kesden:2006vz,Kesden:2006zb,Warnick:2008tk,Peebles:2009th}},
or from caustics of DM \cite{Sanderson:2010ct}, in this case with
a distinct halo collapse threshold  between baryons and coupled matter.
Finally, it is possible to discriminate between potentials in their
response to the amount of coupling. Those behaviours certainly are
calling for further investigations.


\noindent \acknowledgments
The authors would like to thank Nelson Nunes for discussions. TB wishes
to thank Orfeu Bertolami. MLeD also wishes to thank David Mota, for
discussions and transmission of a MATLAB code and José Pedro Mimoso
for many years of support through this project. It also involved the
support CSIC (Spain) under the contract JAEDoc072, with partial support
from CICYT project FPA2006-05807, at the IFT, Universidad Autonoma
de Madrid, Spain. The work of M.Le~D. is now supported by FAPESP
(Brazil), with process number 2011/24089-5, at the DFMA/IF, Universidade
de S\~ao Paulo, Brazil.

\appendix

\section{Unequal initial overdensities\label{sec:Unequal-initial-overdensities}}

In section   \ref{sub:Simplified-evolution}, we assumed that the
initial values for the overdensities in coupled and uncoupled matter
were equal, as in and most works in the field, like \cite{Lokas:2000cn,Mainini:2002ry,Mainini:2003uf,Mota:2004pa,Lokas:2003cj,Solevi:2004tk,Solevi:2005ta,Nunes:2004wn,Manera:2005ct,LeDelliou:2005ig,Tarrant:2011qe}.
It is however possible to keep independent the overdensity of each
species. In particular, the coupled to uncoupled overdensity relation
reads in general 
\begin{align}
1+\delta_{c} & =\left(1+\delta_{u}\right)\frac{1+\delta_{ci}}{1+\delta_{ui}}\frac{e^{B_{\star}-B}}{e^{B_{\star i}-B_{i}}}=\left(1+\delta_{u}\right)\Delta e^{B_{\star}-B},\label{eq:uneqInitialDeltas}
\end{align}
 defining $\Delta=\dfrac{1+\delta_{ci}}{\left(1+\delta_{ui}\right)e^{B_{\star i}-B_{i}}}$,
such that the assumptions of equal matter ($\delta_{ci}=\delta_{ui}$)
and zero quintessence initial overdensities ($\phi_{\star i}=\phi_{i}$)
reduce to $\Delta=1$, and thus eq. (\ref{eq:uneqInitialDeltas})
simplifies into relation (\ref{deltac+t}-a) used in section  \ref{sub:Simplified-evolution}.
For the total matter relation (\ref{deltac+t}-b), the freedom transcribes
as 
\begin{align}
1+\delta_{m}=(1+\delta_{u})\frac{\rho_{u}}{\rho_{m}}+\left(1+\delta_{c}\right)\frac{\rho_{c}}{\rho_{m}} & =\frac{1+\delta_{u}}{\rho_{m}}\left(\rho_{u}+\rho_{c}\Delta e^{B_{\star}-B}\right)\nonumber \\
 & =(1+\delta_{u})\cdot\frac{\Omega_{c0}\,\Delta e^{B_{\star}-B_{0}}+\Omega_{u0}}{\Omega_{c0}\, e^{B-B_{0}}+\Omega_{u0}},\label{eq:totalMatterUneqInit}
\end{align}
 which again simplifies when $\Delta=1$. The linear versions of eqs.
(\ref{eq:uneqInitialDeltas}) and (\ref{eq:totalMatterUneqInit})
then read as 
\begin{align}
\delta_{c,L} & =\delta_{u,L}+B^{\prime}\delta\phi+\delta_{ci}-\delta_{ui}-B_{i}^{\prime}\delta\phi_{i}, & \delta_{m,L} & =\delta_{u,L}+\left(\frac{B^{\prime}\delta\phi+\delta_{ci}-\delta_{ui}-B_{i}^{\prime}\delta\phi_{i}}{1+\frac{\Omega_{u0}}{\Omega_{c0}}\, e^{B_{0}-B}}\right).
\end{align}
 The coupled overdensity equation can then be used, together with
its non-linear counterparts, in the correspondingly modified scheme
formed after eqs. (\ref{eq:KleinGordonCollapse}), (\ref{eq:densityEvolL}),
(\ref{eq:KGlinear}) and (\ref{eq:densityEvolL}) to follow the uncoupled
evolution. Then, the coupled values can be transcribed algebraically
from the uncoupled values. Since $\delta_{ci}$ and $\delta_{ui}$
are small, $\Delta\simeq1$. As the differences in evolution are small
and do not bring further light into the question of the segregated
spherical collapse, we only present in the main paper the results
for $\Delta=1$.

\section{Observational constraints\label{sec:observations}}

\subsection{Supernovae}

We use the updated Union2.1 compilation from the Supernova Cosmology
Project, consisting of a sample of 580 SNeIa \cite{Suzuki:2011hu}.
We constrain our model through fitting the distance modulus $\mu(z)$,
\begin{equation}
\mu(z)=5\log_{10}\big(d_{L}(z)\big)+\mu_{0}
\end{equation}
 where $\mu_{0}=M_{B}-5\log_{10}h$, with $M_{B}$ representing the
absolute magnitude of the SNeIa and $h$ being the Hubble parameter
at present in units of 100 km s$^{-1}$ Mpc$^{-1}$.

The luminosity distance $d_{L}(z)$ can be computed through 
\begin{equation}
d_{L}(z)=(1+z)\int_{0}^{z}\frac{H_{0}}{H(z')}dz'\label{lumdist}
\end{equation}

The $\chi^{2}$ to fit the data is then given by 
\begin{equation}
\chi_{SN}^{2}=\sum_{j=1}^{580}\frac{\big(\mu(z_{j})-\mu_{obs,j}\big)^{2}}{\sigma_{\mu,j}}
\end{equation}
 with $\mu_{obs,j}$ and $\sigma_{\mu,j}$ being the observed distance
modulus and its error for each supernova. We then marginalize this
over the absolute magnitude $M_{B}$ to obtain our final results (see
for example \cite{DiPietro:2002cz,Nesseris:2005ur} for details).

\subsection{Cosmic Microwave Background}

A simple but effective CMB constraint for our models can be obtained
from the shift $R$ parameter as given in the WMAP collaboration 7
year results \cite{Komatsu:2010fb}. The shift parameter \cite{Bond:1997wr}
is given by 
\begin{equation}
R=\frac{\sqrt{\Omega_{m0}}}{1+z_{d}}d_{L}(z_{d})
\end{equation}
 with $d_{L}$ being the luminosity distance \ref{lumdist} and $z_{d}$
the redshift at decoupling. We compute the value of $z_{d}$ using
the standard fitting formula from \cite{Hu:1995en}. We also need
the accoustic scale at decoupling, 
\begin{equation}
l_{A}=\pi\frac{d_{L}(z_{d})}{(1+z_{d})r_{s}(z_{d})}
\end{equation}
 where $r_{s}$ is the comoving sound horizon size at decoupling,
\begin{equation}
r_{s}(z_{d})=\int_{z_{d}}^{\infty}\frac{1}{H(z)\sqrt{3+\frac{9\rho_{b}}{4\rho_{\gamma}}}}dz
\end{equation}
 with $\rho_{b}$ and $\rho_{\gamma}$ being respectively the baryons
and photons energy density.

The $\chi^{2}$ is then given by 
\begin{equation}
\chi_{CMB}^{2}=(\vec{x}-\vec{x}_{obs})^{T}C^{-1}(\vec{x}-\vec{x}_{obs})
\end{equation}
 where $\vec{x}=(l_{A},R,z_{d})$ and $C^{-1}$ is the inverse covariance
matrix from the observations. From WMAP7 we get 
\begin{equation}
\vec{x}_{obs}=(302.09;1.725;1091.3)
\end{equation}
 and 
\begin{equation}
C^{-1}=\begin{bmatrix}2.305 & 29.698 & -1.333\\
29.698 & 6825.270 & -113.180\\
-1.333 & -113.180 & 3.414
\end{bmatrix}
\end{equation}

\subsection{Results}

\begin{figure}
\begin{centering}
\includegraphics[width=0.4\columnwidth]{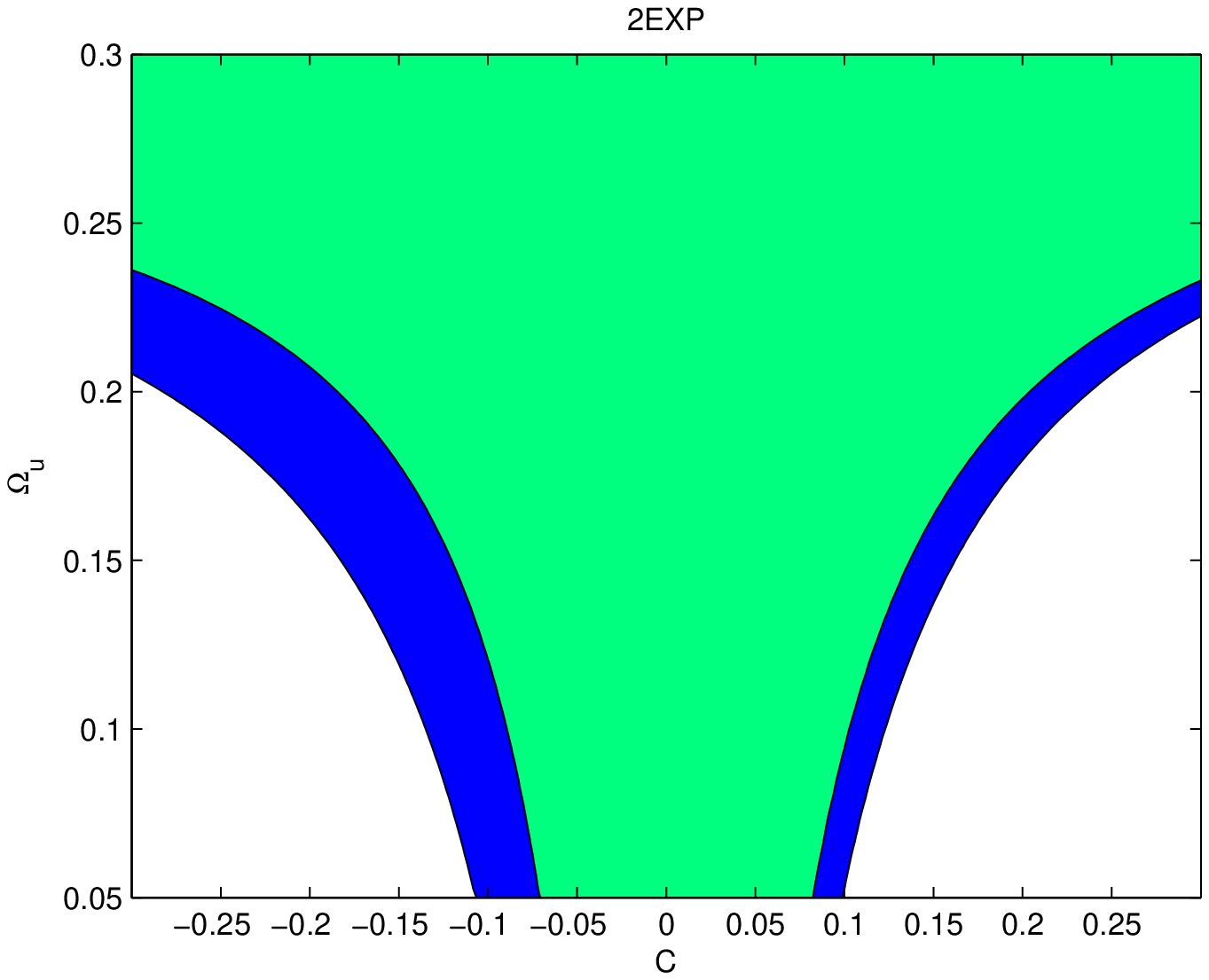} \qquad{}\includegraphics[width=0.4\columnwidth]{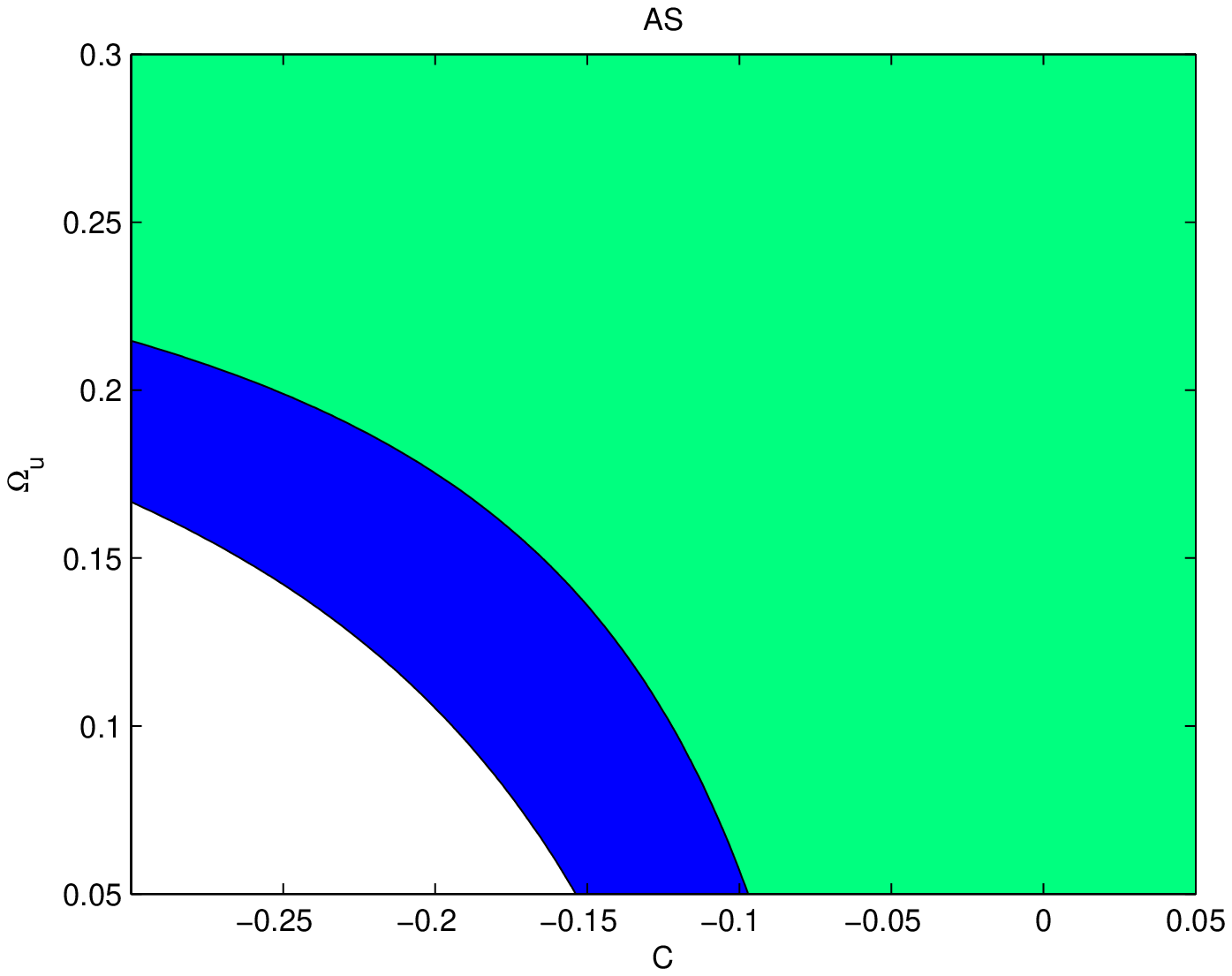}\\
 \includegraphics[width=0.4\columnwidth]{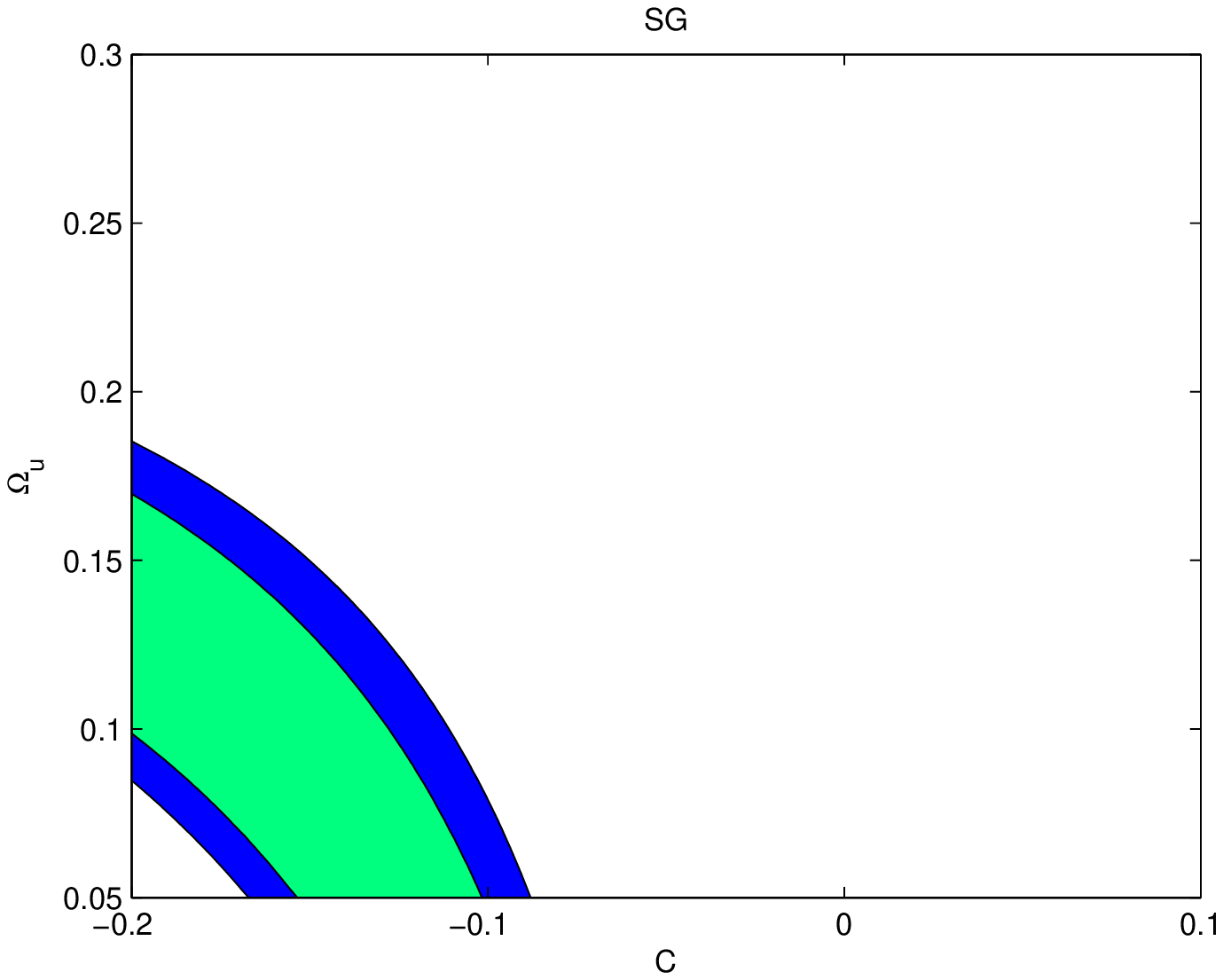} \qquad{}\includegraphics[width=0.4\columnwidth]{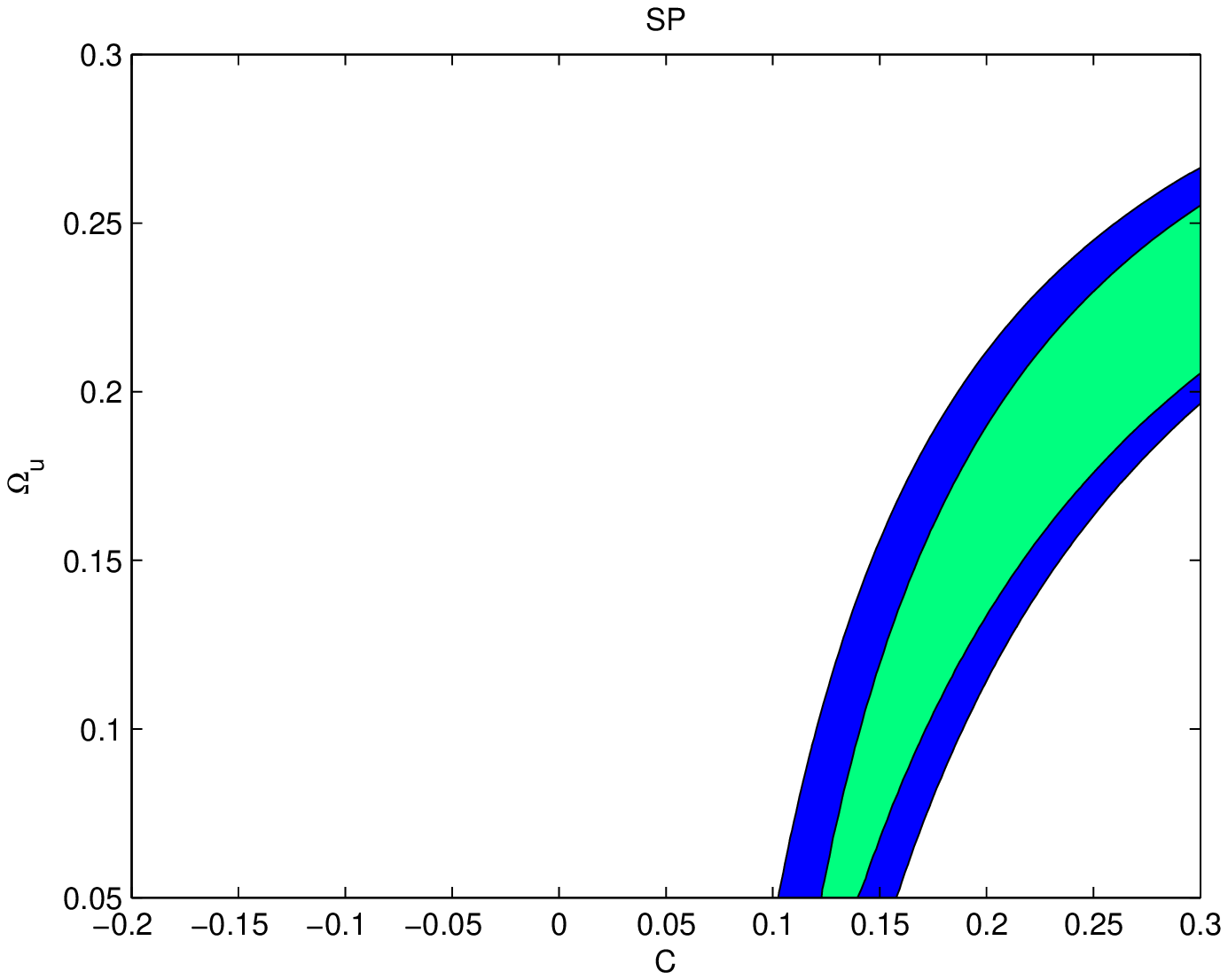}
\par\end{centering}

\caption{\label{obsconst} Two dimensional $\chi^{2}$ likelihood for the double
exponential (2EXP), Albrecht-Skordis (AS), SUGRA (SG) and Steinhardt
(SP) potentials. We show the $68\%$ and $90\%$ confidence levels.}
\end{figure}

We fixed all our parameters except the value of the coupling $C$
and the amount of uncoupled dark matter $\Omega_{u}$. That is, we
fix the amount of coupled dark matter to be $\Omega_{c}=0.3-\Omega_{u}$,
where $\Omega_{u}=\Omega_{b}+\Omega_{udm}$ includes both the baryons
and the uncoupled dark matter components. The results for both constraints
can be seen in Figure~\ref{obsconst}. We can see that present data
favors negative values of $C$ for the SUGRA potential and positive
values of $C$ for the Steinhardt potential, ruling out a scenario
of uncoupled dark matter for these potentials. Consider, however,
that these are just indicative results for the values of the parameters
we use in this paper, and is not a full scale fit of the cosmological
parameters for these models.

Fixing the value of $\Omega_{u}=0.05$, we obtain as the best fit
for the coupling $C$ the values $C=0.01\pm0.05$, $C=0.01_{-0.04}^{+0.06}$,
$C=-0.13\pm0.02$ and $C=0.13\pm0.02$ for the double exponential,
Albert-Skordis, SUGRA and Steinhardt potentials respectively.

\bibliographystyle{JHEP}
\bibliography{baryonDM1}

\end{document}